\documentclass[10pt,superscriptaddress,american,aps,pra,twocolumn,floatfix]{revtex4-1}

\usepackage[usenames,dvipsnames]{color}
\usepackage[latin9]{inputenc}
\usepackage{babel}
\usepackage{float}
\usepackage[caption=false]{subfig}
\usepackage{textcomp}
\usepackage{amsmath,amssymb}
\usepackage{graphicx}
\usepackage{grffile}
\usepackage[normalem]{ulem}
\definecolor{dblue}{rgb}{0,0,.5}
\usepackage[unicode=true,
 bookmarks=true,bookmarksnumbered=false,bookmarksopen=false,
 breaklinks=false,pdfborder={0 0 1},backref=false,colorlinks=true,citecolor=dblue,linkcolor=dblue,urlcolor=dblue]
 {hyperref}
\usepackage[all]{hypcap}

\usepackage{graphicx}
\newcommand{\pUp}{\ensuremath{
  \mathchoice{\vcenter{\hbox{\includegraphics[height=2ex]{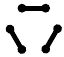}}}}
    {\vcenter{\hbox{\includegraphics[height=2ex]{fig001}}}}
    {\vcenter{\hbox{\includegraphics[height=1.5ex]{fig001}}}}
    {\vcenter{\hbox{\includegraphics[height=1ex]{fig001}}}}
}}
\newcommand{\pDown}{\ensuremath{
  \mathchoice{\vcenter{\hbox{\includegraphics[height=2ex]{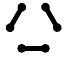}}}}
    {\vcenter{\hbox{\includegraphics[height=2ex]{fig002}}}}
    {\vcenter{\hbox{\includegraphics[height=1.5ex]{fig002}}}}
    {\vcenter{\hbox{includegraphics[height=1ex]{dimerDown}}}}
}}

\newcommand{\bra}[1]{\left\langle #1 \right|}
\newcommand{\ket}[1]{\left| #1 \right\rangle }
\newcommand{\braket}[2]{\left\langle #1 | #2 \right\rangle }
\newcommand{\ketbra}[2]{\left|#1\right\rangle \left\langle #2\right|}

\newcommand{\up}{\uparrow}
\newcommand{\down}{\downarrow}
\newcommand{\VtC}{(V/t)_c}

\newcommand{\QDM}{quantum dimer model}
\newcommand{\QIM}{quantum Ising model}
\newcommand{\CIM}{classical Ising model}

\newcommand{\qdm}{{\mathrm{QDM}}}
\newcommand{\qim}{{\mathrm{QIM}}}
\newcommand{\cim}{{\mathrm{CIM}}}

\newcommand{\eff}{{\mathrm{eff}}}
\newcommand{\Star}  {{\mathrm{star}}}
\newcommand{\Plaq}  {{\mathrm{plaq}}}
\newcommand{\Tr}{\operatorname{Tr}}
\newcommand{\hS}{\hat{\sigma}}
\newcommand{\hm}{\hat{m}}
\newcommand{\hn}{\hat{n}}
\newcommand{\hrho}{\hat{\rho}}
\newcommand{\hB}{\hat{B}}
\renewcommand{\vec}[1]{{\boldsymbol{#1}}}
\newcommand{\vS}{{\vec{\sigma}}}
\newcommand{\mc}[1]{\mathcal{#1}}
\newcommand{\stack}[2]{\genfrac{}{}{0pt}{}{#1}{#2}}

\begin{document}

\title{Phase diagram of the hexagonal lattice quantum dimer model:\texorpdfstring{\\}{} Order parameters, ground-state energy, and gaps}
\newcommand{\jussieu}{Laboratoire de Physique Th\'{e}orique de la Mati\`{e}re Condens\'{e}e, UMR CNRS/Universit\'{e} Pierre et Marie Curie/Sorbonne Universit\'{e}s, 4 Place Jussieu, 75252 Paris Cedex 05, France}
\author{Thiago M. Schlittler} 
\affiliation{\jussieu}
\author{R\'{e}my Mosseri} 
\affiliation{\jussieu}
\author{Thomas Barthel} 
\affiliation{Department of Physics, Duke University, Durham, North Carolina 27708, USA}
\affiliation{Laboratoire de Physique Th\'{e}orique et Mod\`{e}les Statistiques, Universit\'{e} Paris-Sud, CNRS UMR 8626, 91405 Orsay Cedex, France}

\begin{abstract}
The phase diagram of the quantum dimer model on the hexagonal (honeycomb) lattice is computed numerically, extending on earlier work by Moessner, Sondhi, and Chandra. The different ground state phases are studied in detail using several local and global observables. In addition, we analyze imaginary-time correlation functions to determine ground state energies as well as gaps to the first excited states. This leads in particular to a confirmation that the intermediary so-called plaquette phase is gapped -- a point which was previously advocated with general arguments and some data for an order parameter, but required a more direct proof. On the technical side, we describe an efficient world-line quantum Monte Carlo algorithm with improved cluster updates that increase acceptance probabilities by taking account of potential terms of the Hamiltonian during the cluster construction. The Monte Carlo simulations are supplemented with variational computations.
\end{abstract}

\date{August 20, 2017}


\maketitle

\section{Introduction}\label{sec:intro}\vspace{-0.1em}
Interacting spin systems in two dimensions have been widely studied over the last decades, both from experimental and theoretical points of view. Of importance in this context is the so-called resonating valence bond approach put forward by P.\ W.\ Anderson in 1973 \cite{Anderson1973153} in order to analyze the physics of spin 1/2 Heisenberg antiferromagnets. This has later been advocated as a way to study the yet unsolved problem of high-temperature superconductivity. Following Rokhsar and Kivelson \cite{rokhsar_superconductivity_1988}, it proves interesting, when studying the low energy properties of these phases, to consider a simpler model, called the quantum dimer model (QDM). In the latter, the $SU(2)$ singlet bonds are replaced by hard core dimers defined on the edges of the lattice. Quantum dimer models have been employed to study for example superconductivity \cite{rokhsar_superconductivity_1988,Sachdev2000}, frustrated magnets \cite{Moessner2001-86,moessner_ising_2001,Misguich2002-89,moessner_quantum_2011,albuquerque_phase_2011}, or hardcore bosons \cite{SenthilPRB.76.235107}. They can feature topological order, spin liquid phases, and deconfined fractional excitations \cite{moessner_quantum_2011}.

Before enlarging on the quantum systems, let us say a few words about the classical case. Lattice dimer coverings -- the basis states of the Hilbert space in the quantum case -- represent already a rich mathematical problem with many connections to statistical physics problems. For a graph defined by its vertices and edges (defining faces, often called plaquettes in the present context), a dimer covering is a decoration of the bonds, such that every vertex is reached by exactly one dimer. The simplest rearrangement mechanism for dimer coverings is provided by so-called plaquette flips. These are applicable for plaquettes around which every second bond has a dimer and the flip amounts to exchanging covered and uncovered bonds, yielding a different valid dimer covering (e.g., $\pUp\longleftrightarrow \pDown$ for a hexagonal lattice).
Dimer coverings are closely related to other configurational problems. For the hexagonal lattice, these are ground-state configurations of a classical Ising-spin model with antiferromagnetic interactions on the (dual) triangular lattice, planar rhombus tilings, and height models \cite{blotehilhorst_82,elser_solution_1984}.
Topological sectors can be characterized by so-called fluxes (see below). These sectors are invariant under the plaquette flips. The topological properties depend for example on the boundary conditions and have consequences on the physics of the \QDM{}.

The quantum version, as proposed by Rokhsar and Kivelson, corresponds to considering the set of all dimer coverings of the classical problem as an orthonormal basis spanning the Hilbert space. The Hamiltonian contains kinetic terms that correspond precisely to the elementary flips described above and an additional potential term, proportional to the number of flippable plaquettes. The competition between these kinetic and potential terms leads to a non-trivial phase diagram. For example, when the potential term dominates in amplitude and is of negative sign, the ground state is expected to be dominated by configurations which maximize the number of flippable plaquettes; for the opposite sign, one expects a ground state dominated by dimer configurations without flippable plaquettes. As will be discussed, such configurations exist and correspond to the so-called star and staggered phases, respectively. In between these two extremes, the phase diagram can display intermediary phases. The ground state is known exactly for the point where kinetic and potential terms are of equal strength. The physics around this so-called Rokhsar-Kivelson (RK) point is expected to be different for bipartite and non-bipartite lattices \cite{moessner_quantum_2011}.
\begin{figure*}[t]
\begin{centering}
	\subfloat[]
	{
		\includegraphics[width=0.2\textwidth]{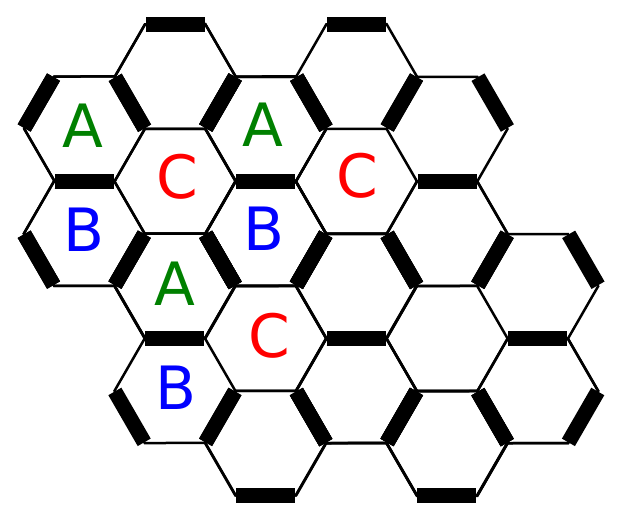}\label{fig:dimer_coverings_star3}
	}\quad\quad
	\subfloat[]
	{
		\includegraphics[width=0.2\textwidth]{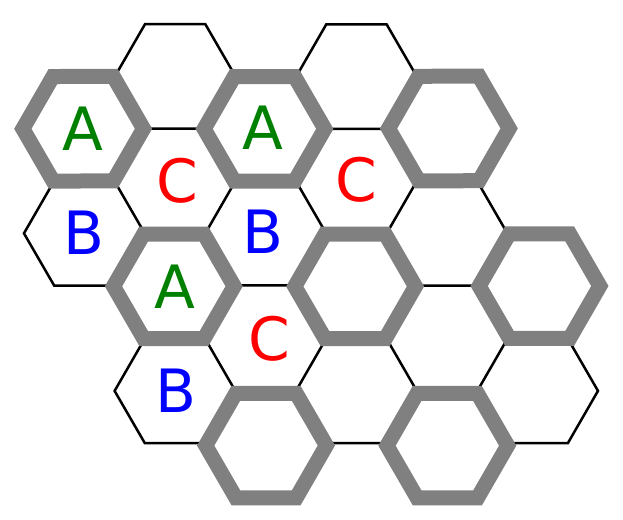}\label{fig:dimer_coverings_plaq}
	}\quad\quad
	\subfloat[]
	{
		\includegraphics[width=0.2\textwidth]{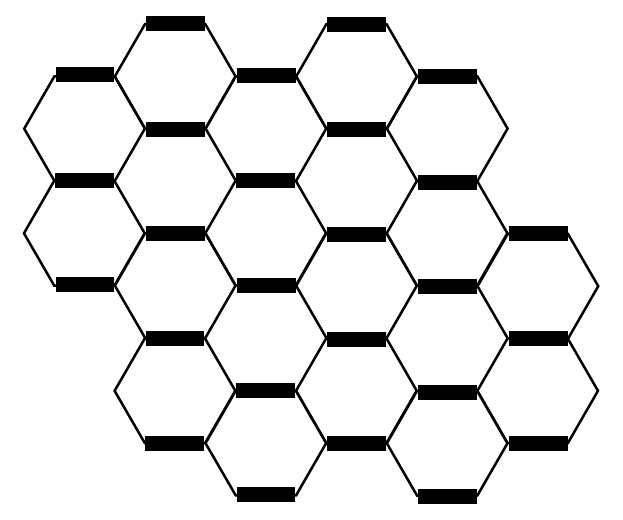}\label{fig:dimer_coverings_stag}
	}
\end{centering}
\caption{Prototypes of quantum dimer ground states on a honeycomb lattice: (a) star phase, (b) plaquette phase, (c) staggered phase. Edges with a high probability of carrying a dimer are indicated in black, and edges with a $\sim 50\%$ probability are indicated in gray. The (dual) lattice can be decomposed into three triangular sublattices $A,B$, and $C$ as shown. In the star state, flippable plaquettes occupy two of the sublattices, while, in the plaquette phase, all plaquettes of one of the sublattices are in a benzene-like resonating state (gray hexagons).}\label{fig:examples-of-dimer}
\end{figure*}

In this paper, we provide an extensive study of the \QDM{} on the bipartite hexagonal (honeycomb) lattice along the lines already followed by Moessner, Sondhi, and Chandra \cite{moessner_phase_2001}. In their seminal work, these authors numerically investigated the phase diagram by studying a local order parameter which, in addition to the generic RK transition point, shows a first order transition which separates the star phase from an intermediary phase, the so-called plaquette phase. See Fig.~\ref{fig:examples-of-dimer} for a sketch of these phases. Based on finite-size data for the star-phase order parameter at three different temperatures, Moessner \emph{et al.}\ argued that the plaquette phase should be gapped -- a point which conflicts with an earlier analytical analysis \cite{orland_exact_1993}. In the present paper, we use quantum Monte Carlo simulations to extend the numerical work by studying order parameters for different system sizes and temperatures as well as ground-state energies and excitation gaps which we obtain from imaginary-time correlation functions. This leads to a clear confirmation of the gapped nature of the plaquette phase. We shortly explain the reason for conflicting results of Ref.~\cite{orland_exact_1993} and supplement the Monte Carlo results with a variational treatment.

The outline of this paper is the following. In section~\ref{sec:qdm}, the quantum dimer Hamiltonian is detailed and the nature of the different phases is explained. In section~\ref{sec:mapping}, we describe the employed world-line quantum Monte Carlo algorithm which is based on a mapping of the two-dimensional (2D) quantum model to a 3D classical problem, and which we accelerate through suitable cluster updates. Section~\ref{sec:observables} introduces the employed observables. In section~\ref{sec:results}, we present the results of the numerical simulations, and characterize the different phases and phase transitions on the basis of different observables, ground-state energies, and energy gaps. Supplementary variational computations are described in section~\ref{sec:variation}. Section~\ref{sec:summary} gives a summary of the results. Detailed discussions of some technical issues are delegated to the appendices.

\section{Quantum Dimer Model}\label{sec:qdm}
\subsection{Hilbert space and Hamiltonian}
We consider the 2D hexagonal lattice of spins-$1/2$ with periodic boundary conditions. As described in the introduction, the \QDM{}s are defined on the subspace spanned by dimer configurations where every spin forms a singlet $(\ket{\up,\down}-\ket{\down,\up})/\sqrt{2}$ with one of its three nearest neighbors. These different dimer configurations are used as an orthonormal Hilbert space basis. Models of this type are for example important in the context of resonating valence bond states and superconductivity \cite{rokhsar_superconductivity_1988,Moessner2001-86,Misguich2002-89}. Note that different dimer coverings of the lattice (dimer product states) are not orthogonal with respect to the conventional inner product for spin-$1/2$ systems ($\braket{\sigma}{\sigma'}=\delta_{\sigma\sigma'}$). However, as  explained in Ref.~\cite{rokhsar_superconductivity_1988}, the two inner products can be related to one another through additional longer-ranged terms in the Hamiltonian that turn out to be not essential. The Hamiltonian
\begin{align}
	\hat{H}_\qdm = &-t\sum_i\left(\ketbra{\pUp_i}{\pDown_i}+h.c.\right)\nonumber\\
	           &+V\sum_i \left(\ketbra{\pUp_i}{\pUp_i}+\ketbra{\pDown_i}{\pDown_i}\right)
	\label{eq:H_QDM}
\end{align}
contains a kinetic term $\propto t$ that flips flippable plaquettes (those with three dimers along the six plaquette edges) and a potential term $\propto V$ that counts the number of flippable plaquettes. The sums in Eq.\ \eqref{eq:H_QDM} run over all plaquettes $i$ of the hexagonal lattice on a torus. The potential term favors ($V<0$) or disfavors ($V>0$) flippable plaquettes. The only free parameter of this model is hence the ratio $V/t$. In Ref.~\cite{Schlittler2015-115}, we discuss a generalized version of the model with additional potential terms.
In the following, a plaquette carrying $j$ dimers is called a $j$-plaquette such that $3$-plaquettes are the flippable ones.

The configuration space of the system is not simply connected but consists of different topological sectors which are not flip-connected. Each sector is characterized by two flux quantum numbers, also known as winding numbers:
Call $\mc{A}$ and $\mc{B}$ the two triangular sublattices of the hexagonal lattice such that all nearest neighbors of any site from $\mc{A}$ are in $\mc{B}$.
To compute the flux $W$ through a cut $\mc{C}$ of the lattice, first orient all cut edges, say, from $\mc{A}$ to $\mc{B}$, weight them by $+2$ or $-1$, depending on whether they are covered by a dimer or not, and multiply each weight by $\pm 1$ according to the orientation of the edge with respect to $\mc{C}$. The flux $W$ is then computed by summing the contributions of all cut edges. Such fluxes $W$ are invariant under plaquette flips. As fluxes through closed contractible curves $\mc{C}$ are zero, one has two flux quantum numbers $W_x$ and $W_y$, corresponding to the two topologically distinct closed non-contractible curves on the torus. Notice that these two fluxes characterize an average slope in the height representation \cite{blotehilhorst_82} of the system.

Let us briefly recall the phase diagram obtained in Ref.~\cite{moessner_phase_2001}. Three phases belonging to two different topological sectors have been described. The ground states for the so-called star phase ($-\infty<V/t<\VtC$) and the plaquette phase ($\VtC<V/t<1$) are found in the zero flux sector, while the staggered phase ground states ($1<V/t<\infty$) are in the highest flux sector. See Fig.~\ref{fig:examples-of-dimer}. The ground states in the zero flux sector can be distinguished using sublattice dimer densities. For that purpose, we recall that the plaquettes of the hexagonal lattice can be separated into three subsets -- triangular sublattices $A$, $B$, and $C$ of disjoint plaquettes, as depicted in Fig.~\ref{fig:examples-of-dimer}, such that every hexagon of a set shares bonds with three hexagons of the two other sets each.

\section{Quantum-classical mapping and Monte Carlo simulation}\label{sec:mapping}
As done by Moessner \emph{et al.}\ \cite{moessner_two-dimensional_2000,moessner_ising_2001,moessner_phase_2001}, the 2D \QDM{} on a hexagonal lattice can be studied by mapping it first to a 2D \QIM{} on the (dual) triangular lattice. The resulting Ising-type quantum model can be studied efficiently using world-line quantum Monte Carlo \cite{Suzuki1976,Suzuki1977-58} by approximating its partition function and observables by those of a classical 3D Ising-type model (CIM) on a stack of triangular 2D lattices (quantum-classical mapping) as described in the following subsections. We accelerate the Monte Carlo simulation of the classical 3D model through suitable cluster updates.

\subsection{Equivalence to a quantum Ising model on the dual lattice}\label{sec:QDM_to_QIM}
\begin{figure}[b]
\begin{centering}
\includegraphics[width=0.95\columnwidth]{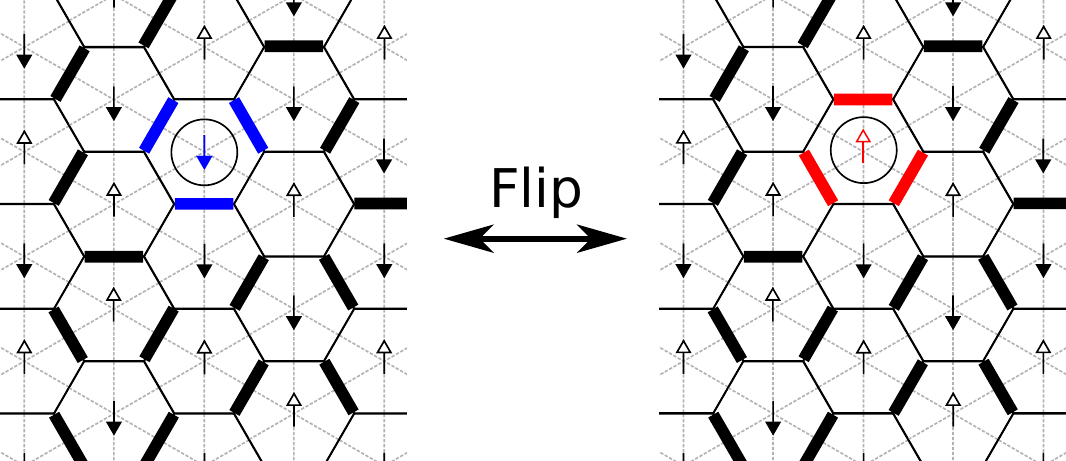}
\end{centering}
\caption{Equivalence of dimer coverings of the hexagonal lattice and Ising-spin configurations on the (dual) triangular lattice. Every dimer corresponds to a frustrated bond ($\up-\up$ or $\down-\down$). Flipping a plaquette in the hexagonal lattice is equivalent to flipping a spin in the dual lattice.}\label{fig:hex_trig}
\end{figure}

As shown in Fig.~\ref{fig:hex_trig}, the dual of the hexagonal lattice is the triangular lattice whose vertices are located at the hexagon centers. We assign a spin-$1/2$ ($\sigma_i=\pm 1$) to each of the vertices and, as explained in the following, the \QDM{} \eqref{eq:H_QDM} maps for the limit $J_z\to\infty$ to the Ising-type quantum model
\begin{equation}\label{eq:H_QIM}
	\hat{H}_\qim = J_z\sum_{\langle i,j\rangle} \hS^z_i\hS^z_j - t\sum_i \hS_i^x + V\sum_i  \delta_{\hat{B}_i,0}
\end{equation}
on the triangular lattice, where $\{\hS_i^x,\hS_i^y,\hS_i^z\}$ denote the Pauli spin matrices for lattice site $i$. The operator $\hat{B}_i:= \sum_{j\in \mc N_i} \hS^z_j$, with $\mc N_i$ being the set of the six nearest neighbors of site $i$, yields for an $\{\hS^z_i\}$-eigenstate the value zero if exactly three of the six bonds starting at site $i$ are frustrated. A bond is called frustrated if the corresponding two spins are parallel.

At the center of each triangle lies a vertex of the hexagonal lattice. For a given dimer covering, one dimer is shared by this vertex and the dimer crosses exactly one of the three edges of the triangle at an angle of 90\textdegree. See Fig.~\ref{fig:hex_trig}. For sufficiently strong $J_z$, the physics of the \QIM{} \eqref{eq:H_QIM} is restricted to the subspace spanned by the classical ground states. Those have exactly one frustrated bond per triangle (all other configurations have higher energy). The identification of dimer basis states and Ising basis states is then straightforward. Given a certain dimer configuration, put a spin up on an arbitrary site. Associating frustrated Ising bonds with those that are crossed by a dimer in the given state, we can work inward-out, assigning further Ising spins until the triangular lattice is filled. The state, up or down, for a new site depends on the spin state of an already assigned neighboring site and on whether the corresponding bond is frustrated or not.
Periodic boundary conditions in the \QDM{} correspond, in the \QIM{}, to periodic boundary conditions in $x$-direction ($y$-direction) for even flux quantum numbers $W_x$ ($W_y$) and to anti-periodic boundary conditions for odd $W_x$ ($W_y$).

This mapping of dimer configurations on the hexagonal lattice to spin-$1/2$ configurations on the triangular lattice implies that, for the \QIM{}, we employ the conventional inner product $\braket{\sigma}{\sigma'}=\delta_{\sigma\sigma'}$ for which different $\{\hS^z_i\}$-eigenstates are orthonormal. In the Hamiltonian \eqref{eq:H_QIM}, the spin-flip terms $\propto t$ correspond to the kinetic term in the \QDM{} \eqref{eq:H_QDM}. Due to the energetic constraint imposed by $J_z\to \infty$, they are only effective for sites where the spin flip does not change the number of frustrated bonds, corresponding to the flippable plaquettes in the dimer model. The term $\propto V$ corresponds exactly to the potential term in the dimer model.

In fact, the mapping of dimer configurations to Ising-spin configurations, described above is 1 to 2, as we are free to choose the orientation of the first assigned spin. This can be fixed by associating the quantum dimer Hilbert space with the spin-flip symmetric subsector in the Ising-spin Hilbert space, i.e., with $\operatorname{span}\{\ket{\vS}+\ket{-\vS}\}$ for all ground states $\vS$ of the classical Ising model. However, for simplicity, the quantum Monte Carlo simulation of the system (see section~\ref{sec:MC}) operates in the full Hilbert space $\operatorname{span}\{\ket{\vS}\}$. This is unproblematic as we are only interested in ground-state properties, and the global ground state resides in the symmetric sector \cite{moessner_ising_2001}:
As all off-diagonal matrix elements of the Ising Hamiltonian \eqref{eq:H_QIM} are non-positive, the ground state can be chosen as a superposition of basis states with non-negative real coefficients according to the Perron-Frobenius theorem. There can only be one such ``nodeless'' energy eigenstate. The restriction of the \QIM{} to the symmetric sector yields a matrix for which, in the basis of states $\ket{\vS}+\ket{-\vS}$, again all off-diagonal elements are non-positive. So, the ground state of this sector is also nodeless and is identical to the global ground state.

\subsection{Approximation by a classical 3D Ising model}\label{sec:QIM_to_CIM}
To apply world-line quantum Monte Carlo \cite{Suzuki1977-58}, we can approximate the partition function and observables of the \QIM{} \eqref{eq:H_QIM} on the 2D triangular lattice by those of a 3D \CIM{} on a stack of 2D triangular lattices by a Trotter-Suzuki decomposition \cite{Trotter1959,Suzuki1976}. To this purpose, we separate the Hamiltonian \eqref{eq:H_QIM} into two parts
\begin{gather*}
	\hat{H}_\qim = \hat{H}^z + \hat{H}^x\quad\text{with}\quad \hat{H}^x :=-t\sum_i\hS^x_i\quad\text{and}\\
	\hat{H}^z :=H^z(\{\hS^z_i\}):=J_z\sum_{\langle i,j\rangle} \hS^z_i\hS^z_j + V\sum_i  \delta_{\hB_i,0}.
\end{gather*}
As detailed in appendix~\ref{appx:QIM_to_CIM}, one can use the Trotter-Suzuki decomposition
\begin{equation*}
	e^{-\beta \hat{H}_\qim}=\left(e^{-\frac{\Delta\beta}{2} \hat{H}^z}e^{-\Delta\beta\hat{H}^x}e^{-\frac{\Delta\beta}{2} \hat{H}^z}\right)^N+\mc{O}(N\Delta\beta^3)
\end{equation*}
of the density operator with imaginary-time step $\Delta\beta\equiv\beta/N$ to determine the parameters $K^z$ and $K^\tau$ for the \CIM{}
\begin{equation}\label{eq:H_CIM}
	E_\cim(\vS) = K^z \sum_{n} H^z(\vS^n)- K^\tau \sum_{n,i} \sigma_i^n\sigma_i^{n+1}
\end{equation}
such that the partition functions $Z_\qim\equiv\Tr e^{-\beta \hat{H}_\qim}$ and $Z_\cim = \sum_\vS e^{-E_\cim(\vS)}$ of the two models coincide (up to a known constant $\mc{A}$),
\begin{subequations}\label{eq:equal_Z}
\begin{align}
	&Z_\qim = \mc{A}\cdot Z_\cim + \mc{O}(N\Delta\beta^3)\\
	&\text{with}\quad \mc{A}=[{\sinh(2\Delta\beta t)}/{2}]^{LN/2}. \label{eq:factor_A}
\end{align}
\end{subequations}
Similarly, for expectation values of observables $\hat O=O(\{\hS^z_i\})$ that are diagonal in the $\{\hS^z_i\}$-eigenbasis,
\begin{subequations}\label{eq:equal_obs}
\begin{align}
	\langle\hat{O}\rangle_\qim &{=} \langle O\rangle_\cim + \mc{O}(N\Delta\beta^3),\quad\text{where}\\
	\langle\hat{O}\rangle_\qim &\equiv \frac{1}{Z_\qim}{\Tr( e^{-\beta \hat{H}}\hat{O})}\quad\text{and} \label{eq:diagObs_QIM}\\
	\langle O\rangle_\cim &\equiv \frac{1}{Z_\cim}{\sum_\vS e^{-E_\cim(\vS)} O(\vS^n)}\quad\forall_n. \label{eq:diagObs_CIM}
\end{align}
\end{subequations}
In these equations, $\vS=(\vS^n|n=1,\dots,N)$ is a vector of classical ground-state spin configurations $\vS^n=(\sigma^n_i| i\in\mc T)$ on the triangular lattice $\mc T$ for each of the imaginary-time slices $n=1,\dots,N$, and $L$ is the number of lattice sites $i$ in $\mc T$.

As shown in appendix~\ref{appx:QIM_to_CIM}, the parameters $K^z$ and $K^\tau$ of the \CIM{} \eqref{eq:H_CIM} are given by
\begin{equation}\label{eq:param_connection}
	K^z=\Delta\beta \quad\text{and}\quad
	e^{-2K^\tau}=\tanh(\Delta\beta t).
\end{equation}

Besides computing in this way expectation values of diagonal operators $\hat O=O(\{\hS^z_i\})$, one can also evaluate expectation values of non-diagonal operators like the energy expectation value of the \QDM{} by evaluating corresponding imaginary-time correlation functions in the \CIM{}. See appendix~\ref{appx:energy}.

\subsection{Monte Carlo algorithm with cluster updates}\label{sec:MC}
The representation \eqref{eq:diagObs_CIM} of expectation values \eqref{eq:diagObs_QIM} of quantum observables as expectation values of classical observables is of great value, as it can be evaluated efficiently with a Monte Carlo algorithm by sampling classical states $\vS$. Specifically, one generates a Markov chain of classical states $\vS$ with probabilities $e^{-E_\cim(\vS)}/Z_\cim$ and averages $O(\vS)$ over these states. 

The most simple scheme would be to choose in every iteration of the algorithm one of the flippable spins (a spin on site $j$ of time slice $n$ is flippable, iff $\sum_{i\in\mc{N}_j}\sigma_i^n=0$), compute the energy difference $E_\cim(\vS')-E_\cim(\vS)$ that the flipping of the spin would cause, and flip it with a probability that is given by the so-called Metropolis rule as detailed in appendix~\ref{appx:MC}.

However, as one increases the accuracy by reducing $\Delta\beta$ (for a fixed inverse temperature $\beta=N\Delta\beta$), the coupling $K^{\tau}$ of the time slices increases with $K^\tau\propto \log (1/\Delta\beta t)$ and the \CIM{}, hence, becomes stiff with respect to the time direction. In the generated states $\vS$, there will occur larger and larger 1D clusters of spins along the time-direction that have the same orientation, $\sigma_i^{m}=\sigma_i^{m+1}=\dots=\sigma_i^{m+n}$. Flipping one of the spins inside such a cluster becomes less and less frequent as the associated energy change increases with the increasing coupling $K^\tau$. This would result in an inefficient Monte Carlo sampling with high rejection rates. We avoid this effect by doing 1D cluster updates instead of single-spin updates: In every iteration of the algorithm, an initial flippable spin is selected and, in an intermediate phase, a 1D cluster is grown in the imaginary-time direction before suggesting to flip this cluster as a whole. We further decrease rejection rates by taking account of changes in the number of flippable spins during the cluster construction. See appendix~\ref{appx:MC} for details.

\section{Studied observables}\label{sec:observables}
In the next section, section \ref{sec:results}, we numerically characterize the phase diagram of the \QDM{} using several observables: the magnetization of the associated Ising model, dimer densities, the ground-state energy, and the energy gap to the first excited state. Let us briefly describe them in the following.

\subsection{Magnetization}\label{sec:m_RMS}
We compute the root mean square (RMS) $z$ magnetization $\langle \hm_z^2\rangle^{1/2}:=\langle(\sum_i\hS^z_i/L)^2\rangle_\qim^{1/2}$ for the \QIM{} \eqref{eq:H_QIM} with $L$ sites. As explained below, $\langle \hm_z^2\rangle^{1/2}=1/3$ for $V/t\to-\infty$ and $\langle \hm_z^2\rangle^{1/2}=0$ for $V/t>1$, such that the RMS $z$ magnetization is an order parameter. It also facilitates comparison to earlier work \cite{moessner_phase_2001}. In the thermodynamic limit, $\langle \hm_z^2\rangle^{1/2}$ vanishes in the plaquette phase and can hence be used to locate the transition from the star phase to the plaquette phase. We also compute the $x$ magnetization $\langle\hm_x\rangle:=\langle\sum_i\hS^x_i/L\rangle_\qim$, which corresponds to the kinetic energy and helps to assert the finite gap in the plaquette phase.

\subsection{Local and global dimer observables}
The simulations give access to dimer densities $\langle \hn_i\rangle$, the average number of dimers on plaquette $i$. Two-dimensional (contrast) plots of these densities nicely illustrate the ground-state structure.

We also evaluate the normalized total numbers of $j$-plaquettes $(\langle \hrho_0\rangle,\langle \hrho_1\rangle,\langle \hrho_2\rangle,\langle \hrho_3\rangle)$. Specifically, with $j$-plaquettes being the plaquettes carrying $j$ dimers and $L$ being the system size, $\hrho_j\equiv\sum_i \delta_{\hn_i,j}/ L$. As described in appendix~\ref{appx:sumrule}, the plaquette numbers $\langle\hrho_j\rangle$ obey the sum rule
\begin{equation}\label{eq:sumrule}
	\langle \hrho_3 \rangle - \langle \hrho_1 \rangle - 2\langle \hrho_0 \rangle = 0. 	
\end{equation}
Notice that $\langle\hrho_2\rangle$ does not enter in the sum rule, while changes in the number of 3-plaquettes, which enter both the kinetic and potential energy terms, must be compensated by plaquettes with zero dimers or one dimer. $\langle \hrho_2 \rangle$ is nevertheless constrained by the fact the total number of plaquettes is of course constant, i.e., $\sum_{j=0}^3 \langle\hrho_j\rangle= 1$.

\subsection{Sublattice dimer densities}\label{sec:sublattices}
As described above and indicated in Fig.~\ref{fig:examples-of-dimer}, the hexagonal plaquettes can be separated into three sets $(A,B,C)$, each forming a triangular lattice, such that every hexagon in a set shares a bond with three hexagons of the two other sublattices each. The ``prototype'' states of the star and the plaquette phases (Fig.~\ref{fig:examples-of-dimer}) can be characterized qualitatively in terms of dimer densities in the three sublattices. To this purpose, we can analyze averaged dimer densities on each sublattice and call them $\langle \hn_{A,B,C}\rangle$, i.e., $\hn_A\equiv \frac{3}{L} \sum_{i\in A} \hn_i$, etc.

It should be stressed that the systems under study may have degenerate (or nearly degenerate) ground states. The star crystal (ground state for $V/t=-\infty$) and the ideal plaquette state (not a ground state, see below) are both threefold degenerate. For sufficiently large systems, it is expected that this symmetry is kinetically broken in the Monte Carlo simulation. However, one cannot fully prevent the system from translating from one typical ground-state configuration to another (even at the level of medium-size patches), smearing out the information carried by these local parameters. This possibility was minimized here by choosing large system sizes and low temperatures. We nevertheless carefully kept track of this possible problem in analyzing the data. Specifically for the sublattice dimer densities, during the course of the Monte Carlo simulation, we have rearranged the three sublattice labels according to the dimer occupancies, instead of keeping the ordering constant.
\begin{figure*}[t]
\centering
	\includegraphics[width=0.95\textwidth]{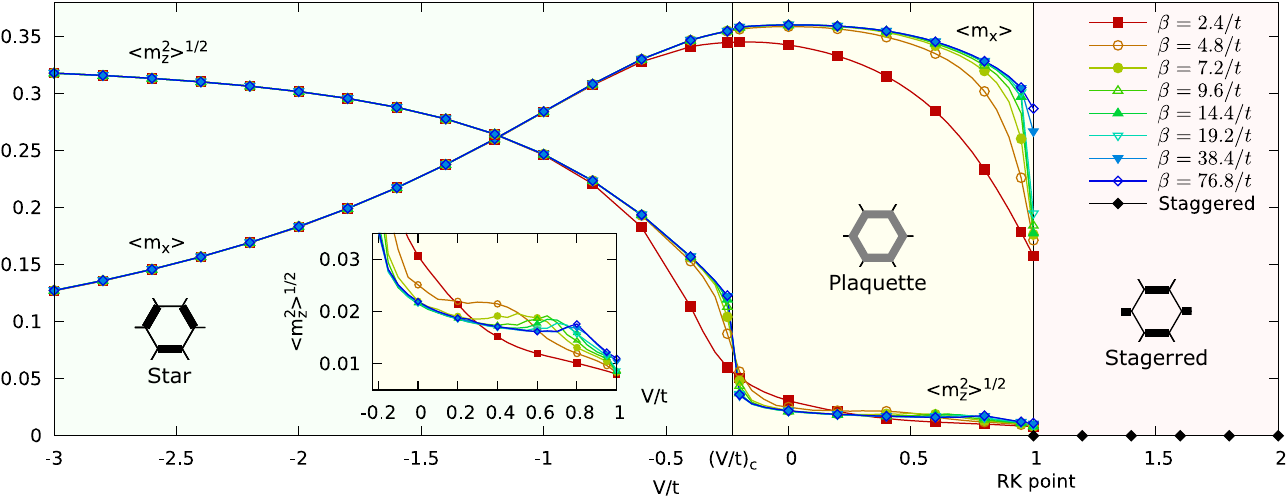}
\caption{The root-mean square $z$ magnetization $\langle \hm_z^2\rangle^{1/2}$ and the $x$ magnetization $\langle \hm_x\rangle$ for the \QDM, as defined in section~\ref{sec:m_RMS}. The different curves correspond to different inverse temperatures $\beta$ and are obtained from Monte-Carlo simulations for $V/t \leq 1$ with system size $L=36\times 36$ and $\Delta\beta=0.02/t$. For all $V/t>1$, the staggered state, depicted in Fig.~\ref{fig:dimer_coverings_stag}, is the ground state and hence $\langle \hm_z^2\rangle^{1/2},\langle\hm_x\rangle=0$. To improve visibility, not all data points are indicated by symbols.}\label{fig:mag_vs_vt_large}
\end{figure*}

\subsection{Ground state energy}
To study the phase diagram, it is certainly of high interest to access the ground-state energy which directly decides what phase prevails for given values of the Hamiltonian parameters. For sufficiently low temperatures in the simulation, the expectation value $\langle\hat{H}_\qim\rangle$ of the \QIM{} Hamiltonian corresponds to the ground-state energy. But $\hat{H}_\qim$ is not a diagonal operator, and hence Eq.\ \eqref{eq:equal_obs} cannot be used. As detailed in appendix~\ref{appx:energy}, it can nevertheless be evaluated on the basis of imaginary-time correlators  $\langle\sigma_i^n\sigma_i^{n+1}\rangle_\cim$. 

\subsection{Energy gap}
It is important to determine whether a given phase has gapless excitations or not.
As explained in appendix~\ref{appx:gap}, we can estimate the energy gap to the first excited state by fitting imaginary-time correlation functions $\langle \hat A(0)\hat A^\dag(\tau)\rangle$. In the \CIM{} \eqref{eq:H_CIM} they correspond to inter-layer correlators with layer distance $\Delta n=\tau/\Delta\beta$. For sufficiently low temperatures, and  $\tau$ and $\beta - \tau$ big compared to the gap to the second excited state, the leading terms in the correlation function are of the form $a+b\cdot\cosh( (\beta/2-\tau)  \Delta E)$, allowing to fit the upper bound $\Delta E$ of the gap.

\section{Simulation results}\label{sec:results}
\begin{figure*}[t]
\centering
\subfloat[]{
	\includegraphics[height=0.30\textwidth]{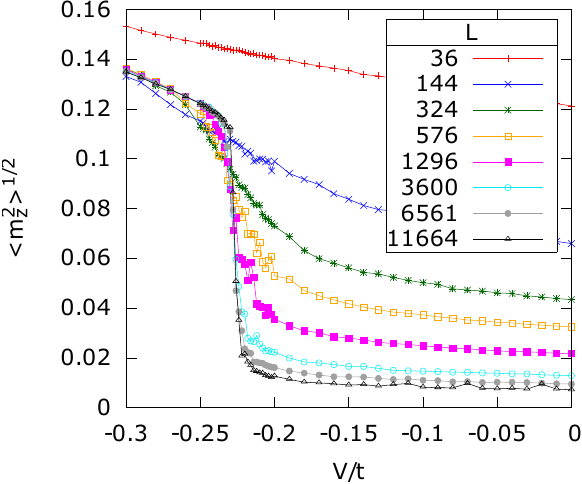}\label{fig:mag_vs_vt}
}
\subfloat[]{
	\includegraphics[height=0.30\textwidth]{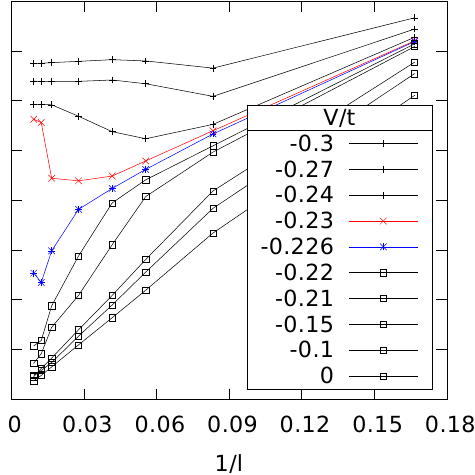}\label{fig:mag_vs_1_L}
}
\subfloat[]{
	\includegraphics[height=0.30\textwidth]{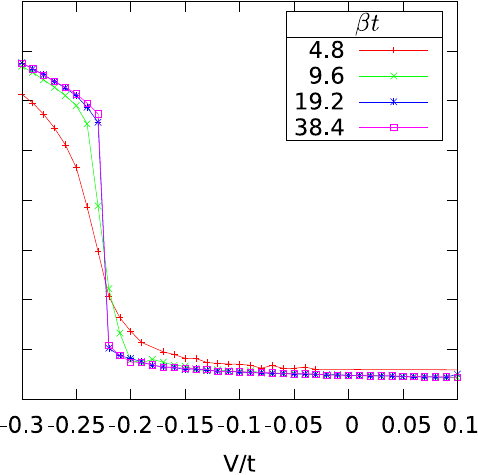}\label{fig:mag_vs_vt_tempN}
}
\caption{Locating the transition between the star and plaquette phases. (a) The root-mean square $z$ magnetization $\langle \hm_z^2\rangle^{1/2}$ for different lattice sizes $L=\ell^2$
as a function of $V/t$ for $\beta=19.2/t$ and $\Delta\beta=0.02/t$. (b) For the same temperature, $\langle \hm_z^2\rangle^{1/2}$ is plotted for different values of $V/t$ as a function of the inverse linear system size $1/\ell$. (c) $\langle \hm_z^2\rangle^{1/2}$ as a function of $V/t$ for different temperatures, $\Delta\beta = 0.02/t$, and $L = 81\times 81$.}
\end{figure*}
In the following, let us study in detail the phase diagram of the \QDM{}, starting from large negative $V/t$, i.e., in the star phase. The observables described in the previous section are evaluated in simulations for patches of linear size $\ell$ with a $60^\circ$ rhombus shape, periodic boundary conditions, and $L=\ell^2$ plaquettes. In order to be able to separate the lattice into the three sublattices $A$, $B$, and $C$ (Section~\ref{sec:sublattices}), $\ell$ needs to be a multiple of three.

\subsection{The star phase (\texorpdfstring{$\boldsymbol{-\infty<V/t<\VtC}$}{-infinity < V/t < (V/t)\_c})}\label{ssec:star_phase}
This phase has previously been called the ``columnar phase'', in analogy with a corresponding phase of the square lattice \QDM{}, where dimers are aligned along columns. For the hexagonal lattice, this denomination is a bit misleading, and we follow Ref.~\cite{JulienPRB.89.201103} in calling it the ``star phase'' \footnote{The name ``star phase'' originates from the rhombus tiling associated to this dimer configuration in the limit $V\to -\infty$, which is either known as the ``star tiling'' or the ``dice lattice''.}. 

For large negative $V$, the potential term dominates the kinetic term and the ground state is dominated by dimer configurations that maximize the number of flippable plaquettes. In the limit $V\to -\infty$, there are three degenerate ground states given by ideal star states as depicted in Fig.~\ref{fig:dimer_coverings_star3}, where all plaquettes from two of the three sublattices, say $A$ and $B$, are flippable, while all plaquettes of the third sublattice ($C$) are dimer-free.
\begin{equation}\label{eq:starState}
	\ket{\psi_\Star} = \bigotimes_{i\in A}\ket{\pUp_i} \bigotimes_{j\in B} \ket{\pDown_j}
\end{equation}
Changing from the dimer to the Ising-spin representation, we have
\begin{equation*}
	\ket{\psi_\Star} = \frac{1}{\sqrt{2}}\Big(\bigotimes_{i\in A\cup B}\ket{\up_i} \bigotimes_{j\in C} \ket{\down_j} + \up\leftrightarrow\down\Big).
\end{equation*}
$A$ and $B$ carry spins of equal orientation, and all spins on sublattice $C$ have the opposite orientation such that the RMS $z$ magnetization reaches its maximum possible value $\langle \hm_z^2\rangle^{1/2}=1/3$. In the real system, it will decrease as $V/t$ is increased. Figure~\ref{fig:mag_vs_vt_large} shows that $\langle \hm_z^2\rangle^{1/2}$ is still very close to the maximum value $1/3$ at $V/t = -3$. The $x$ magnetization $\langle \hm_x\rangle$ is zero in the ideal star state $\ket{\psi_\Star}$, hence it vanishes for $V/t\to -\infty$ in the real system.
\begin{figure}[t]
\begin{centering}
\includegraphics[width=0.9\columnwidth]{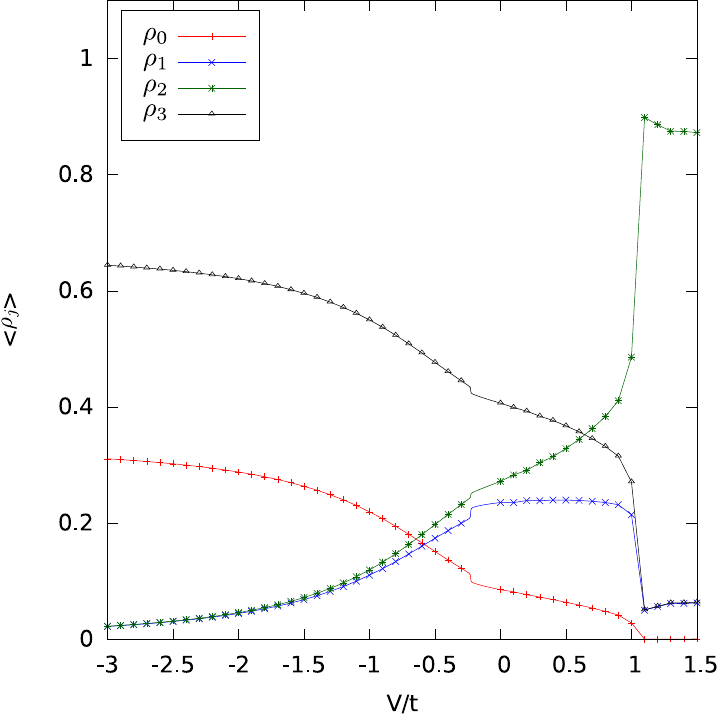}
\end{centering}
\caption{Normalized numbers of $j$-plaquettes, $\langle\hrho_j\rangle$, for the zero flux sector, system size $L = 81\times 81$, $\beta=19.2/t$, and $\Delta\beta=0.02/t$. Around $\VtC$, a finer grid of points was used to resolve the jumps in the densities at the transition. In that region, data points are not marked by symbols. Although the global ground state is not in the zero flux sector for $V/t>1$, data obtained for the zero flux sector is also shown for that region and is discussed in the text.}\label{fig:n_i}
\end{figure}
\begin{figure*}[t]
	\centering
	\subfloat[]{
		\includegraphics[width=0.95\columnwidth]{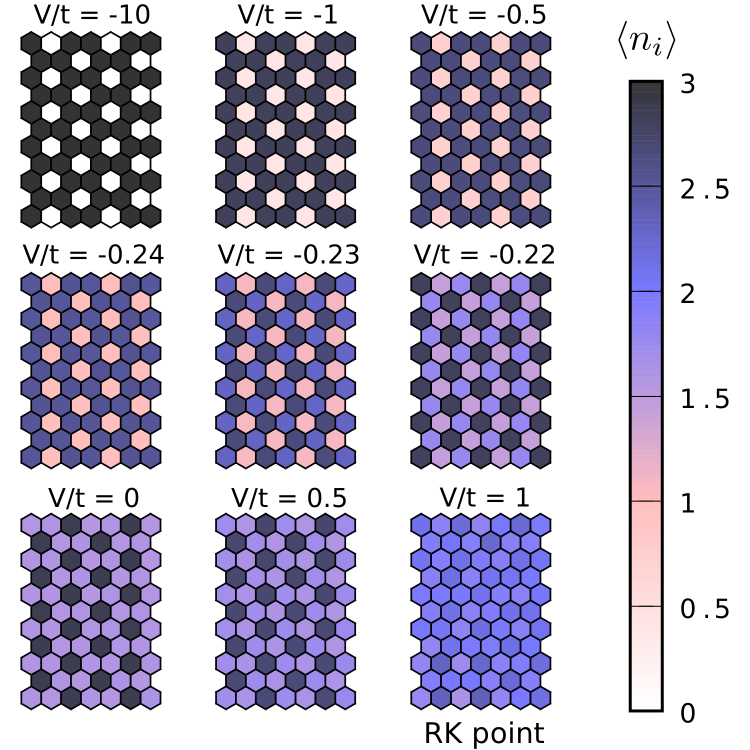}\label{fig:dimer_per_site}
	}
	\subfloat[]{
		\includegraphics[width=0.95\columnwidth]{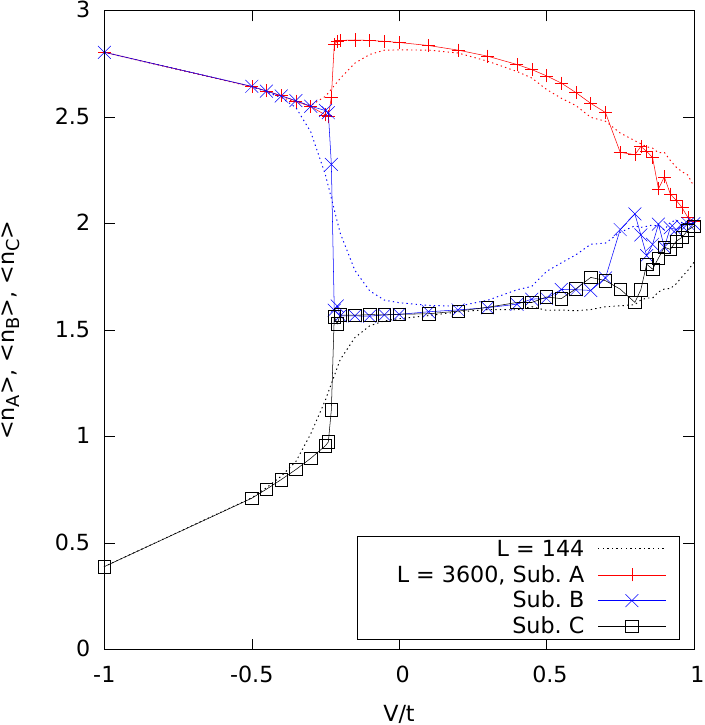}\label{fig:sublattice_vs_vt}}
\caption{(a) Local dimer density $\langle\hn_i\rangle$ for different values of $V/t$ with $L=60\times 60$ plaquettes, $\beta=19.2/t$, and $\Delta\beta=0.02/t$. (b) Sublattice dimer densities $\langle \hn_{A,B,C}\rangle$ as functions of $V/t$ for $L=60\times 60$ (solid lines) and $L=12\times 12$ plaquettes (dashed lines), respectively.}\label{fig:dimers_substruct}
\end{figure*}
   
To understand how increasing $V/t$ affects the star phase ground state, one can do perturbation theory in $t/V$. The calculation, done up to second order in $t/V$, is given in appendix~\ref{appx:pert}. The result for the ground-state energy is shown in Fig.~\ref{fig:q_energy_pert}. It compares well with the simulation results up to $V/t\sim -1$. The first correction to the ideal star state amounts to mixing in configurations with one flipped plaquette. 
 
The ground state for small negative $t/V$ is the ideal star state dressed with flipped plaquettes in both $A$ and $B$, and, at some point also in $C$, when three or more flips have occurred locally. These changes in the ground sate can be quantified by the numbers of $j$-plaquettes as done in Fig.~\ref{fig:n_i}. In the ideal star state ($V/t\to-\infty$), one has $(\langle \hrho_0\rangle,\langle \hrho_1\rangle,\langle \hrho_2\rangle,\langle \hrho_3\rangle)=(1/3,0,0,2/3)$. Say, sublattices $A$ and $B$ contain the flippable plaquettes in this limit. After flipping a plaquette in $A$, the three neighboring plaquettes in sublattice $B$ carry two instead of three dimers and the three neighboring plaquettes in $C$ are no more dimer free, but carry one dimer each. The numbers of 0- and 3-plaquettes are hence reduced by three and those of 1- and 2-plaquettes are increased by 3. This explains why the curves for $\langle\hrho_0\rangle$ and $\langle\hrho_3\rangle$ in Fig.~\ref{fig:n_i} are almost parallel up to $V/t\sim -1$ and why those for $\langle\hrho_1\rangle$ and $\langle\hrho_2\rangle$ increase correspondingly and are on top of each other. The contrast plots of the dimer density $\langle\hn_i\rangle$ in Fig.~\ref{fig:dimer_per_site} and the plot of the sublattice dimer densities $\langle \hn_{A,B,C}\rangle$ in Fig.~\ref{fig:sublattice_vs_vt} show that the differences between dimer densities on sublattices $A$ and $B$ on one hand and those on sublattice $C$ on the other hand are reduced before reaching a critical point $\VtC$. 
\begin{figure*}[t]
\centering
\subfloat[]{
\includegraphics[width=0.95\columnwidth]{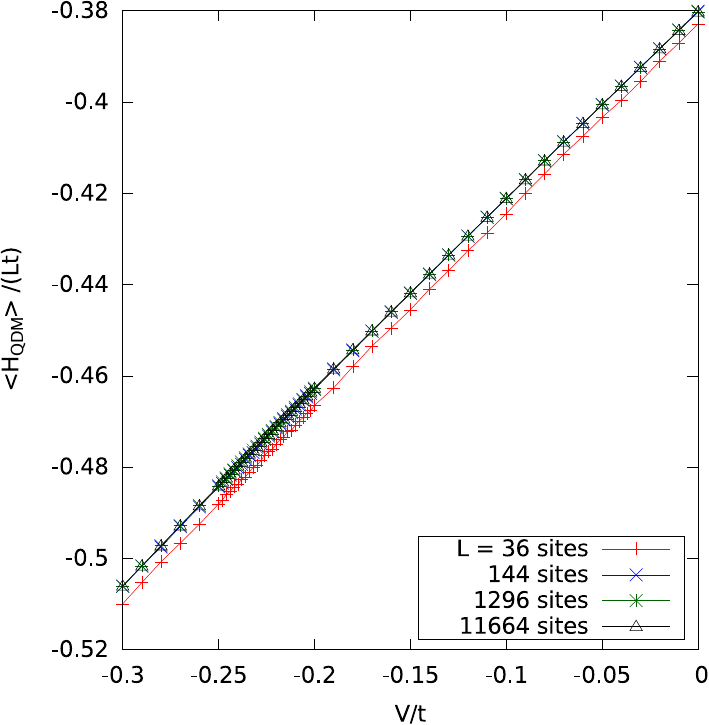}\label{fig:q_energy}
}
\subfloat[]{
\includegraphics[width=0.95\columnwidth]{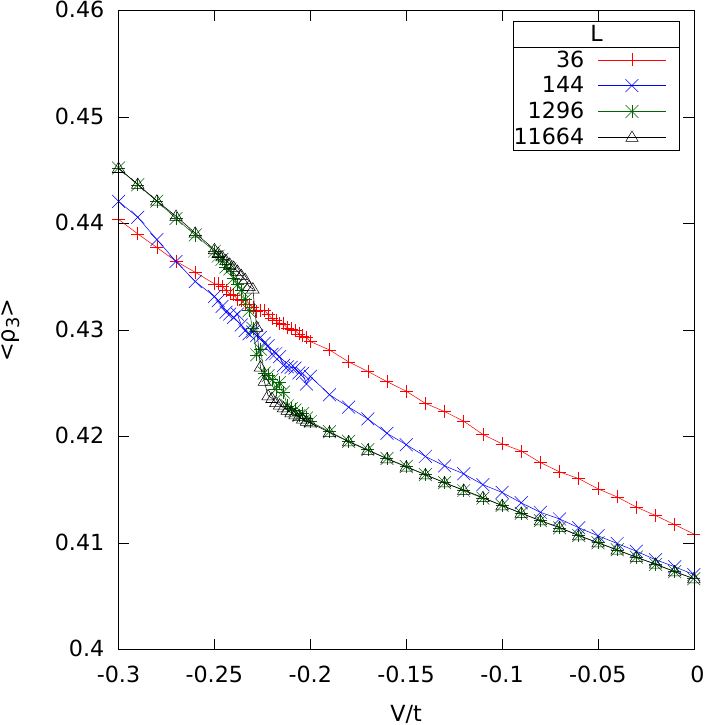}\label{fig:n_3}
}
\caption{The vicinity of the first order phase transition at $\VtC$ for different lattice sizes with $\beta=19.2/t$ and $\Delta\beta=0.02/t$: (a) While the ground-state energy density, $\langle\hat{H}_\qdm\rangle/(Lt)$, is continuous near $\VtC$, (b), the density of flippable 3-plaquettes, $\langle\hrho_3\rangle$, displays a small but evident jump.}
\end{figure*}
 
\subsection{The star to plaquette phase transition at \texorpdfstring{$\boldsymbol{\VtC=-0.228\pm0.002}$}{V/t\_C = -0.228 +/- 0.002}}\label{ssec:star_plaquette}
A first order transition occurring between the star phase and the so-called plaquette phase is found at $\VtC=-0.228\pm0.002$. This critical value is consistent, but more precise than that given in Ref.~\cite{moessner_phase_2001}.
At $\VtC$, the RMS $z$ magnetization $\langle \hm_z^2\rangle^{1/2}$ suddenly drops to a much smaller value which goes to zero in the thermodynamic limit. Figure~\ref{fig:mag_vs_vt_large} displays the RMS $z$ magnetization for the whole phase diagram, while Fig.~\ref{fig:mag_vs_vt} provides a zoom close to the transition and data for different system sizes. We determined $\VtC$ by plotting $\langle \hm_z^2\rangle^{1/2}$ as a function of the inverse (linear) size of the system (Fig.~\ref{fig:mag_vs_1_L}). The temperature dependence shown in Fig.~\ref{fig:mag_vs_vt_tempN} indicates that using a larger $\beta$ should not substantially modify the numerical results. Considering this, we set $\beta = 19.2/t$ and $\Delta\beta = 0.02/t$ for most of our simulations.

The transition can also be observed in the dimer observables. The normalized $j$-plaquette numbers $\langle \hrho_j\rangle$ all show a small discontinuity at $\VtC$ (Fig.~\ref{fig:n_i}). The discontinuity of $\langle \hrho_3\rangle$ (see Fig.~\ref{fig:n_3}) attests the first order character of the transition, since $\langle \hrho_3\rangle$ is the derivative of the energy with respect to $V$. But the discontinuity is quite small leading to a barely visible slope change for the energy (Fig.~\ref{fig:q_energy}).
 
At least as spectacular as the $z$-magnetization drop is the sudden shift in sublattices dimer densities seen in Fig.~\ref{fig:sublattice_vs_vt}. It nicely agrees with the qualitative properties of the ideal plaquette state, depicted in Fig.~\ref{fig:dimer_coverings_plaq}, where the resonating 3-plaquettes are located on one of the three sublattices.

\subsection{The plaquette phase (\texorpdfstring{$\boldsymbol{\VtC<V/t<1}$}{V/t\_C < V/t < 1}) }\label{ssec:plaquette_phase} 
The plaquette phase is more complex to describe than the star phase. The features of the ground state in this phase are to some extent captured by the ``ideal'' resonating plaquette state, a simple product state in which all plaquettes of one of the three sublattices, say $A$, are in a benzene-like resonating state such that
\begin{equation}\label{eq:ideal_plaq}
	\ket{\psi_\Plaq}=\bigotimes_{i\in A} \left(\ket{\pUp_i}+\ket{\pDown_i}\right)/\sqrt{2},
\end{equation}
In contrast to the star phase case, the ideal plaquette state is not an exact ground state for any $V/t$. But the actual ground states are adiabatically connected to the ideal plaquette state.
In the Ising-spin representation, the ideal plaquette state $\ket{\psi_\Plaq}$ reads
\begin{equation*}
	\ket{\psi_\Plaq} = \frac{1}{\sqrt{2}}\bigotimes_{i\in A}\ket{\rightarrow_i} \Big(\bigotimes_{j\in B} \ket{\uparrow_j} \bigotimes_{k\in C} \ket{\downarrow_k}+\up\leftrightarrow\down\Big),
\end{equation*}
where $\ket{\rightarrow_i}$ denotes the $\hS^x_i$-eigenstate $\left(\ket{\uparrow_i}+\ket{\downarrow_i}\right)/\sqrt{2}$. The spins in sublattices $B$ and $C$ must be anti-parallel with respect to each other.
In accordance with the numerical results, the ideal plaquette state has $x$ magnetization $\langle \hm_x\rangle=1/3$ and its RMS $z$ magnetization $\langle \hm_z^2\rangle^{1/2}$ vanishes in the thermodynamical limit: As $\sum_i\hS_i^z\ket{\psi_\Plaq}=\sum_ {i\in A}\hS_i^z\ket{\psi_\Plaq}$, we have that 
\begin{multline}\label{eq:mzScaling}
	\bra{\psi_\Plaq} \hm_z^2\ket{\psi_\Plaq}
	 = \bra{\psi_\Plaq}\Big(\sum_{i\in A}\hS_i^z\Big)^2 \ket{\psi_\Plaq}/L^2\\
	 =\sum_{i\in A}\bra{\psi_\Plaq}(\hS_i^z)^2 \ket{\psi_\Plaq}/L^2 = \frac{1}{3L}\to 0.
\end{multline}
\begin{figure}[t]
\begin{centering}
\includegraphics[width=\columnwidth]{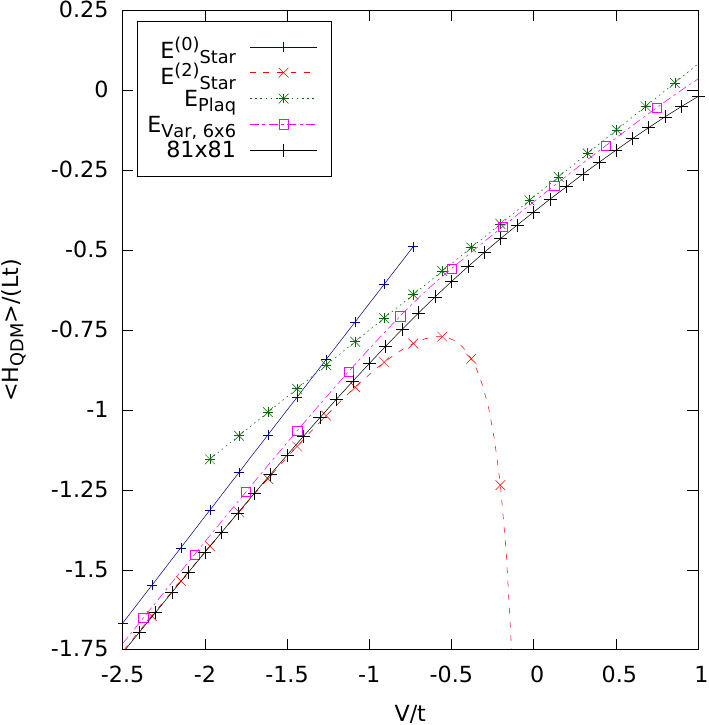}
\end{centering}
\caption{Numerically computed energy density $\langle\hat{H}_\qdm\rangle/L$ for $\beta=19.2/t$, $\Delta\beta=0.02/t$, and $L=81\times 81$, compared to variational and perturbative estimates as described in section~\ref{sec:variation} and appendix~\ref{appx:pert}, respectively.}\label{fig:q_energy_pert}
\end{figure}

The energy density for $\ket{\psi_\Plaq}$ can be computed easily and yields an upper bound to the exact ground state energy. See section~\ref{sec:variation}. At $V=0$, it takes for example the value $-t/3$ which is clearly above the numerically determined value of $\approx -0.38t$ (Fig.~\ref{fig:q_energy}).  One can improve $\ket{\psi_\Plaq}$ as a variational state by adding flip excitations in sublattices $B$ and $C$. This is possible due the fact that $3$-plaquettes occur in $B$ and $C$ with density $1/8$.

\begin{figure}[t]
\begin{centering}
\includegraphics[width=\columnwidth]{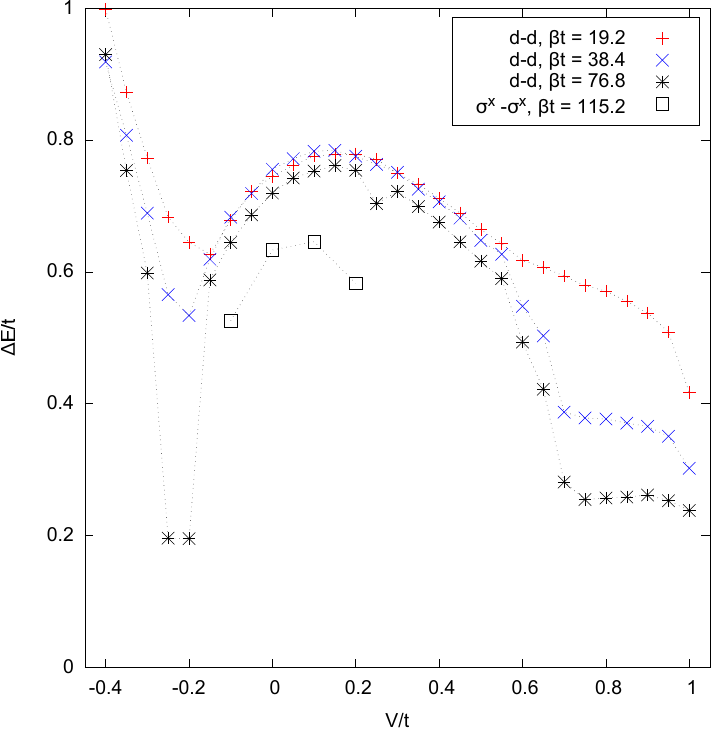}
\end{centering}
\caption{Estimates for the energy gap $\Delta E$ to the first excited state. The gaps were obtained from fits of imaginary-time auto-correlation functions $\langle\hat{d}_{i,j}(0)\hat{d}_{i,j}(\tau)\rangle_\qim$ and $\langle\hS^x_i(0)\hS^x_i(\tau)\rangle_\qim$, for a system with $L=36\times 36$ sites and different temperatures. The results should be interpreted as upper bounds to the real gap, which are close to the actual gap after convergence in $\beta$.}\label{fig:gaps}
\end{figure}

\begin{figure}[t]
\begin{centering}
\includegraphics[width=\columnwidth]{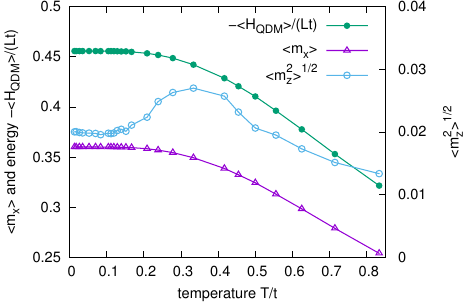}
\end{centering}
\caption{The (exponential) convergence of the energy expectation value $-\langle\hat{H}_\qdm\rangle$ and the magnetizations $\langle \hm_z^2\rangle^{1/2}$ and $\langle \hm_x\rangle$ at finite temperatures, here for $V/t=0.1$ and $L=32\times 32$, confirms that the system is gapped in the plaquette phase.}\label{fig:Tconverge}
\end{figure}

A finite energy gap for the plaquette phase was advocated in Ref.~\cite{Moessner2001-86} with an indirect numerical confirmation based on the observation that RMS magnetizations for three different temperatures coincided at $V/t=0$. Finding this at a single point is not fully conclusive and is also complicated by the fact that the RMS magnetization vanishes in the plaquette phase according to Eq.~\eqref{eq:mzScaling}.
It is possible to estimate excitation gaps, more directly, on the basis of imaginary-time correlation functions. The computation, based on dimer-dimer correlators and plaquette-flip correlators, is described in appendix~\ref{appx:gap}. Our results are presented in Fig.~\ref{fig:gaps}: Starting from the star phase, the gap estimate decreases distinctly around the first order phase transition at $\VtC$. Then, it increases again in the plaquette phase, and eventually goes to zero as we approach the RK point at $V/t=1$. This is clear evidence for a finite gap in the plaquette phase with a value of about $0.6t$ at $V/t=0.1$. This is further supported by the (exponential) convergence of different observables at finite temperatures as shown in Fig.~\ref{fig:Tconverge}.
In the vicinity of the RK point, gap estimates in Fig.~\ref{fig:gaps}, computed from correlators at different temperatures, do not coincide anymore. The reason is simply that, as the gap vanishes, temperatures have to be reduced more and more to obtain the actual gap from the imaginary-time correlators. Also, fitting the correlation functions becomes more difficult as their decay ultimately changes from exponential to algebraic. 

In contrast, an earlier analytical treatment in Ref.~\cite{orland_exact_1993} suggests that the plaquette phase should be gapless. We believe that this is due to a mistake in that derivation. In Ref.~\cite{orland_exact_1993}, the model for $V=0$ and a hexagonal lattice with mixed boundary conditions is mapped to a model of vertically fluctuating non-intersecting strings on a square lattice. See also Ref.~\cite{Schlittler2015-115} for a detailed discussion. First, one can then obtain the ground state of a single string which corresponds to the ground state of the XX chain (energy $E^{(1)}_0\to-2\ell t/\pi$) and that of the \QDM{} in a high-flux sector. One can now add further strings, each reducing the flux by one. To construct an $N$-string ground state, in Ref.~\cite{orland_exact_1993}, the product of vertically shifted single-string ground state wavefunctions is considered. To take account of the no-intersection constraint for the strings, this wavefunction is anti-symmetrized with respect to the string positions, first with respect to all variables $y^{(n)}_1$, then with respect to all $y^{(n)}_2$, etc., where $(x,y^{(n)}_x)$ are the coordinates of string $n$. In analogy to the anti-symmetrization for fermions, it is being assumed in Ref.~\cite{orland_exact_1993} that the resulting state has energy $NE^{(1)}_0$ and is hence the $N$-string ground state. Generalizing the procedure to excited states, gapless excitations are found which simply correspond to gapless excitations of a single string. The described anti-symmetrization, also employed in Refs.~\cite{Orland1992-372,Orland1994-49}, appears to be flawed. Different from the conventional anti-symmetrization for fermions, the resulting $N$-string wavefunction is not a sum of product states but contains also entangled states. Hence, the resulting state is not an energy eigenstate. \footnote{As an example for the conventional anti-symmetrization, consider two non-interacting fermions in 2D space. For a product state $\mu(x^{(1)},y^{(1)})\nu(x^{(2)},y^{(2)})$, one obtains
$\mu(x^{(1)},y^{(1)})\nu(x^{(2)},y^{(2)}) - \mu(x^{(2)},y^{(2)})\nu(x^{(1)},y^{(1)})$. It has zero amplitude for $(x^{(1)},y^{(1)})=(x^{(2)},y^{(2)})$ and has the same energy $E_\mu+E_\nu$ as the original state. For two strings of length two, the anti-symmetrization of Ref.~\cite{orland_exact_1993} yields the state $\mu(y^{(1)}_1,y^{(1)}_2)\nu(y^{(2)}_1,y^{(2)}_2) - \mu(y^{(2)}_1,y^{(1)}_2)\nu(y^{(1)}_1,y^{(2)}_2) - \mu(y^{(1)}_1,y^{(2)}_2)\nu(y^{(2)}_1,y^{(1)}_2) +\mu(y^{(2)}_1,y^{(2)}_2)\nu(y^{(1)}_1,y^{(1)}_2)$. While it is zero for intersecting strings, the second and third components in the sum are not products of single-string states. Hence, the resulting energy is not simply $E_\mu+E_\nu$. Strings interact by restricting their lateral fluctuations. The actual ground state energy per string is hence in general larger than the single-string ground state energy.}

Let us look at further observables to better understand the plaquette phase. The normalized $j$-plaquette numbers $\langle \hrho_j \rangle$ are shown in Fig.~\ref{fig:n_i}. They appear to be much more sensitive to variations in $V/t$ than the RMS magnetization. As $V/t$ increases, $\langle\hrho_3\rangle$ and $\langle\hrho_0\rangle$ continuously decreases while $\langle\hrho_2\rangle$ increases, and $\langle\hrho_1\rangle$ stays almost constant, assuming its maximal value in the phase diagram. The constant and maximal value of $\langle\hrho_1\rangle\approx 0.25$ seems to be a characteristic signature for the plaquette phase. For the ideal plaquette state $\ket{\psi_\Plaq}$, one obtains $(\langle \hrho_0\rangle,\langle \hrho_1\rangle,\langle \hrho_2\rangle,\langle \hrho_3\rangle)=(1/12,1/4,1/4,5/12)$. For no value of $V/t$ do we find agreement with these values, showing once again the difference between the ideal and real plaquette states. Let us now discuss the approach to the RK point.

\subsection{From the plaquette phase to the RK point}\label{ssec:path_to_RK}
The current understanding is that, for bipartite lattices, there occurs a continuous transition from the plaquette phase to the RK point, the latter being an isolated critical point.
Some of the observables, like the dimer densities (Fig.~\ref{fig:n_i}), show indeed the expected smooth behavior. Nevertheless, the RMS $z$-magnetization curves, displayed in Fig.~\ref{fig:mag_vs_vt_large}, show a small feature before the RK point, the $x$ magnetization converges slowly as a function of $\beta$, and sublattice dimer densities in Fig.~\ref{fig:sublattice_vs_vt} show large fluctuations in the interval $0.7<\VtC<1$.

The most natural explanation for this behavior is finite size effects, and the vanishing of the gap in the vicinity of the RK point which leads to an enhancement of fluctuations -- the divergence of dimer-dimer correlation lengths -- and a critical slowing down of the Monte Carlo simulation. The observed effects can be attributed to a crystalline regime with approximate $U(1)$ symmetry in the vicinity of the RK point. The continuum version of the height representation \cite{blotehilhorst_82} of the \QDM{} has $U(1)$ symmetry and algebraically decaying correlations at the RK point $V/t=1$. For $V/t<1$, close to the RK point, there are two length scales, one beyond which dimer-dimer correlators show exponential decay signaling crystalline order, and one beyond which one can observe the breaking of the $U(1)$ symmetry. A linear system size in-between these two length scales corresponds to the crystalline $U(1)$ regime \cite{Fradkin2004-69}.

\subsection{The Rokhsar-Kivelson point (\texorpdfstring{$\boldsymbol{V/t=1}$}{V/t = 1})}\label{ssec:RK_point}
The Rokhsar-Kivelson point is the only point of the phase diagram where the system does not display local order. At this point, the Hamiltonian $\hat{H}_\qdm$ is a sum of projection operators,
\begin{align*}
	\hat{H}_{\qdm,RK} = &-V\sum_i\left(\ketbra{\pUp_i}{\pDown_i}+h.c.\right)\\
	        &\phantom{-}+V\sum_i \left(\ketbra{\pUp_i}{\pUp_i}+\ketbra{\pDown_i}{\pDown_i}\right)\\
		=&V\sum_i\left(\ket{\pUp_i}-\ket{\pDown_i}\right)\cdot\left(\bra{\pUp_i}-\bra{\pDown_i}\right).
\end{align*}
Therefore, the ground-state energy vanishes. For each topological sector, and for each flip-connected subspace in a topological sector, one can build a zero-energy state as an equal-amplitude superposition of all dimer coverings in the corresponding basis set. 

At the RK point, many physical properties, like dimer-dimer correlations, can be derived from the classical dimer problem at infinite temperature. See for instance Ref.~\cite{henley_classical_2004}, where the relation between \QDM{}s at the RK point and their classical counterparts is discussed. We used this relation to benchmark the QMC simulations and, in Ref.~\cite{Schlittler2015-115}, we used it for a perturbative analysis of an extended \QDM{} in the vicinity of the RK point.

\subsection{Staggered phase (\texorpdfstring{$\boldsymbol{1<V/t<\infty}$}{1 < V/t < infinity})}\label{ssec:energy}
In the parameter region $1<V/t<\infty$, flippable plaquettes are disfavored. As the Hamiltonian becomes a sum of projection operators with positive coefficients,
\begin{multline}
	\hat{H}_{\qdm} = t\sum_i\left(\ket{\pUp_i}-\ket{\pDown_i}\right)\cdot\left(\bra{\pUp_i}-\bra{\pDown_i}\right)\\
	+(V-t)\sum_i \left(\ketbra{\pUp_i}{\pUp_i}+\ketbra{\pDown_i}{\pDown_i}\right),
\end{multline}
the ground state energy is non-negative. The ground states are dynamically isolated states, corresponding to dimer coverings without any flippable plaquettes -- so-called staggered configurations. One such state is shown in Fig.~\ref{fig:examples-of-dimer}c. In the Ising-spin representation, they are $\{\hS^z_i\}$-eigenstates with total magnetization zero such that $\langle\hm_z^2\rangle^{1/2}=0$ and $\langle\hm_x\rangle=0$.

Staggered states belong to the topological sectors of highest flux, are zero-energy eigenstates of $\hat{H}_\qdm$ for all values of $V/t$, and become ground states for $V/t\geq 1$. The transition on the right of the RK point is abrupt. At the RK point, all topological sectors contain (at least) one state of vanishing energy. Only the isolated ground states in the maximum flux sectors persist for $V/t>1$.

In the zero flux sector, the RK point corresponds to a first order transition to states with a large majority of 2-plaquettes ($\langle\hrho_2\rangle>0.8$), vanishing $\langle\hrho_0\rangle$, and finite but small values of $\langle\hrho_1\rangle=\langle\hrho_3\rangle$. See Fig.~\ref{fig:n_i}.

\section{Variational treatment}\label{sec:variation}
\begin{figure*}[t]
\centering
\includegraphics[width=0.94\textwidth]{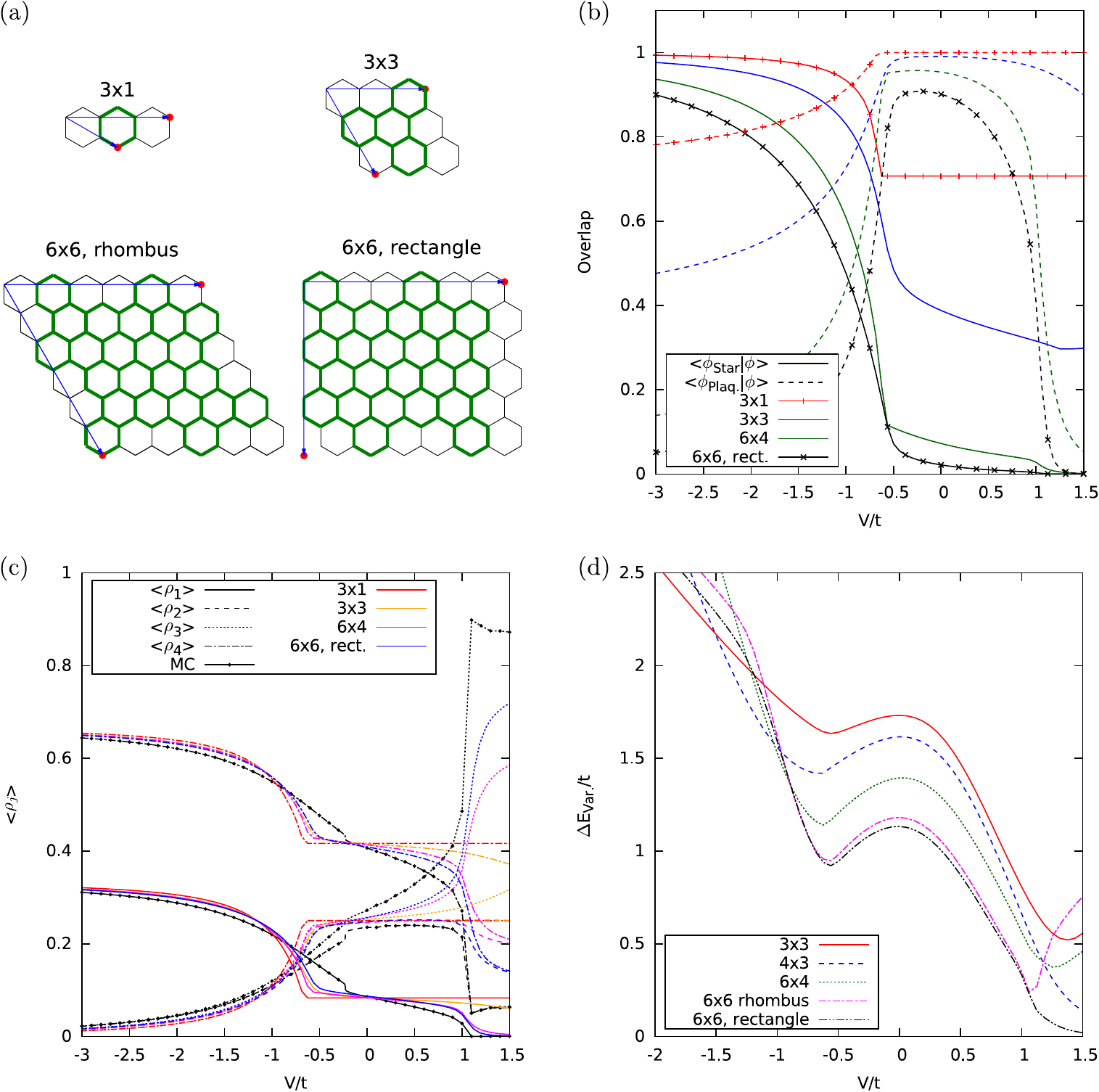}
\caption{Variational treatment for which the energy expectation value for a cell product state $\ket{\psi}=\ket{\phi}\otimes\ket{\phi}\otimes\ket{\phi}\dots$ is minimized with respect to $\ket{\phi}$. (a) Examples for the employed rectangular and lozenge cell shapes. The considered basis states for each cell are all dimer coverings of the marked edges. (b) Overlap of the cell state $\ket{\phi}$ with the ideal star state $\ket{\phi_\Star}$ and the ideal plaquette state $\ket{\phi_\Plaq}$. (c) Normalized numbers of $j$-plaquettes, $\langle\hrho_j\rangle$. (d) Local excitation gap as defined in the text.}\label{fig:var}
\end{figure*}
Let us supplement the Monte Carlo study with a variational treatment. The main motivations are to find states that improve upon the ideal plaquette state to approximate the ground states in the plaquette phase and to obtain further information on excitation gaps.

The ideal plaquette state \eqref{eq:ideal_plaq} is a simple tensor product state with resonating 3-plaquettes on one of the three equivalent sublattices, say sublattice $A$, such that $\ket{\psi_\Plaq}=\bigotimes_{i\in A} \left(\ket{\pUp_i}+\ket{\pDown_i}\right)/\sqrt{2}$.
Recall that $\ket{\psi_\Plaq}$ is not an exact ground state for any value of $V/t$. Its energy expectation value yields hence an upper bound to the ground sate energy. The contribution of the kinetic terms is due to the resonating 3-plaquettes (density $1/3$) and has the value $-tL/3$. The contribution of the potential terms is due to the $L/3$ flippable plaquettes of sublattice $A$, while sublattices $B$ and $C$ contribute with a 3-plaquette density of $1/8$ each. This leads us to 
\begin{equation*}
	E_\Plaq = -\frac{L}{3}t+\left(\frac{L}{3}+\frac{2L}{3}\frac{1}{8}\right)V
	        = L\left(-\frac{1}{3}t + \frac{5}{12}V\right).
\end{equation*}
For $V=0$, this gives an energy of $-t/3$ per plaquette, slightly above the numerically observed value $\approx -0.38t$.

Improving this variational energy is possible along several ways. A simple method is to decompose the lattice into cells as exemplified in Fig.~\ref{fig:var}a and to consider a tensor product
\begin{equation}
	\ket{\Phi}=\ket{\phi}\otimes\ket{\phi}\otimes\ket{\phi}\dots
\end{equation}
of states $\ket{\phi}$ defined on appropriately chosen subgraphs in each cell (bold edges in Fig.~\ref{fig:var}a). We choose these subgraphs to contain all vertices of the $A$-hexagons in the cell and all edges connecting these vertices. The cell Hilbert space is spanned by all dimer coverings of the chosen subgraphs. This construction guarantees that indeed every vertex of the full lattice is reached by exactly one dimer. The cell state $\ket{\phi}$ is determined by minimizing the expectation value of the energy density $\bra{\Phi}\hat{H}_\qdm\ket{\Phi}/L$ with respect to $\ket{\phi}$ under the normalization constraint $\|\phi\|=1$. For the minimization of the energy functional, which is generally a sixth order polynomial in the basis coefficients, we employed the L-BFGS algorithm \cite{Byrd1995-16}, starting from several different initial states to find the global minimum.

The simplest choice is the $3\times 1$ cell depicted in Fig.~\ref{fig:var}a which corresponds to considering states $\ket{\phi}=a\ket{\pUp}_A+b\ket{\pDown}_A$ with $a^2+b^2=1$. The energy functional $-2tab+V(a^2+b^2+a^6+b^6)$ is minimized by
\begin{alignat*}{11}
	&a=-\frac{1}{6}\sqrt{18-6\sqrt{9-4t^2/V^2}}	&\quad\text{for}\quad& V/t&&<&&-2/3\\
	&\text{and by}\quad a=1/\sqrt{2}		&\quad\text{for}\quad& V/t&&\geq&&-2/3,
\end{alignat*}
i.e., for $V/t\geq-2/3$, the solution is given by the ideal plaquette state \eqref{eq:ideal_plaq}. This is reflected in the overlap $\braket{\phi}{\phi_\Star}=1/\sqrt{2}$ to the ideal cell star state and the overlap $\braket{\phi}{\phi_\Plaq}=1$ to the ideal cell plaquette state for $V/t\geq -2/3$ in Fig.~\ref{fig:var}b and the constant normalized numbers of $j$-plaquettes $(\langle \hrho_0\rangle,\langle \hrho_1\rangle,\langle \hrho_2\rangle,\langle \hrho_3\rangle)=(1/12,1/4,1/4,5/12)$ for $V/t\geq -2/3$ in Fig.~\ref{fig:var}c.

When increasing the cell size up to $6\times 6$ rectangles or lozenges, more and more hexagons of the $B$ and $C$ lattices can be flipped, the variational energy density decreases (see Fig.~\ref{fig:q_energy_pert}) and observables such as the $\langle \hrho_i\rangle$ approach the values observed in the Monte Carlo simulations. The overlaps to the ideal star and plaquette states, displayed in Fig.~\ref{fig:var}b, decay with increasing cell size. This is due to two effects. On the one hand, more and more corrections to the ideal states are taken account of and, on the other hand, there is a type of orthogonality catastrophe that is inevitable in the thermodynamic limit. In contrast, fidelities for any fixed-size subregion in the center of the cells would converge. 

The variational treatment can also be used to obtain approximations to the excitation gap. To this purpose we first obtain the optimal cell state $\ket{\phi}$. Singling out a certain cell and fixing state $\ket{\phi}$ on all other cells, we then compute an effective Hamiltonian
\begin{equation*}
	\bra{n}\hat{H}^\mathrm{cell}_\eff\ket{n'} := \big(\bra{n}\otimes\bra{\phi}\otimes\bra{\phi}\dots\big)\hat{H}\big(\ket{n'}\otimes\ket{\phi}\otimes\ket{\phi}\dots\big)
\end{equation*}
for the cell. The gap between the ground state and the first excited state of $\hat{H}^\mathrm{cell}_\eff$, which converges to the gap of $\hat{H}$, is displayed in Fig.~\ref{fig:var}d. It shows the same properties already observed in the Monte Carlo computations (Fig.~\ref{fig:gaps}): a local maximum of the gap inside the plaquette phase region, and a vanishing of the gap in the vicinity of the RK point.

\section{Summary}\label{sec:summary}
We have studied in detail the phase diagram of the quantum dimer model on the hexagonal (honeycomb) lattice \eqref{eq:H_QDM}. To this purpose, we employed world-line quantum Monte Carlo simulations based on approximating the partition function and observables of the 2D quantum system by those of a 3D classical model. We accelerated the algorithm by using improved cluster updates. 

In comparison to earlier work in Ref.~\cite{moessner_phase_2001}, we have used larger systems at lower temperatures to reduce finite-size effects and have investigated several observables in order to give an in-depth description of the ground states and phase transitions. The numbers of $j$-plaquettes and sublattice dimer densities are monitored throughout the phase diagram. In addition, we computed the ground-state energy and energy gaps to the first excited states on the basis of imaginary-time correlation functions. 

The first order transition from the star phase to the plaquette phase is found to occur at $\VtC=-0.228\pm0.002$ and the corresponding symmetry change is clearly reflected in the computed sublattice dimer densities. We also shed some light on the differences between the actual ground states and the corresponding ``ideal'' star and plaquette states, as witnessed by the computed ground-state energies and the behavior of the different dimer observables.

A main result of the present paper is strong numerical evidence for a finite excitation gap in the plaquette phase. At $V/t=0.1$, we find a gap of about $0.6t$, using imaginary-time dimer-dimer correlators and plaquette-flip correlators. This is further supported by the convergence of different observables at finite temperatures. For the attainable cell sizes ($6\times 6$ plaquettes) in the variational treatment, the obtained excitation gap estimate of about $1t$ is still rather large.

In Ref.~\cite{Schlittler2015-115}, we discuss a generalized version of the model \eqref{eq:H_QDM} with additional potential terms. Besides the star plaquette, and staggered phases, it features a plethora further crystalline phases. Their transitions form a fractal structure in the phase diagram, corresponding to a devil's staircase (see also Ref.~\cite{Fradkin2004-69}).

\acknowledgments
We are grateful to G.\ Misguich, J.\ Vidal, and R.\ Moessner for valuable discussions about the quantum dimer problem. G.\ Misguich also provided exact diagonalization data for benchmarking purposes.

\appendix

\section{Sum rule for the plaquette types} \label{appx:sumrule}
Dimer coverings of regular lattices are constrained by simple sum rules, associated to Euler-Poincar\'{e} and Gauss-Bonnet relations for tilings on compact surfaces~\cite{sadocmosseribook99}. 

For a given dimer covering of the hexagonal lattice on a torus, let $N_j$  denote the total number of $j$-plaquettes, i.e., plaquettes covered with $j$ dimers, and let $L$ denote the total number of plaquettes. The $N_j$ obey the sum rules
\begin{equation}\label{eq:sumrules}
 \sum_{j=0}^{3}N_j=L,\quad \sum_{j=0}^{3}jN_j =2L.  
\end{equation}
The first rule signifies that every plaquette can carry from $j=0$ to $j=3$ dimers. The left-hand side of the second rule gives two times the total number of dimers, as every (dimer-carrying) edge belongs to two plaquettes. The right-hand side $2L$ is due to the fact that the total number of vertices is $2L$, and every vertex is reached by exactly one dimer. From the two rules \eqref{eq:sumrules}, it follows that $N_3=N_1+2N_0$ as stated in Eq.~\eqref{eq:sumrule}. Notice that, on average, plaquettes carry two dimers.

\section{From the 2D quantum Ising model to a classical 3D Ising model}\label{appx:QIM_to_CIM}
In Sec.~\ref{sec:QIM_to_CIM}, we have described how the Ising-type quantum model \eqref{eq:H_QIM} on the 2D triangular lattice can be mapped to a 3D \CIM{} on a stack of 2D triangular lattices to allow for an efficient world-line Monte Carlo simulation. Let us show here that the partition functions and expectation values of diagonal observables do indeed coincide up to corrections that are of third order in the imaginary-time step $\Delta\beta=\beta/N$, as claimed in Eqs.\ \eqref{eq:equal_Z} and \eqref{eq:equal_obs}.

Based on the second order Trotter-Suzuki decomposition 
\begin{equation*}
	e^{\lambda(\hat{A}+\hat{B})}=e^{\frac{\lambda}{2}\hat{A}}e^{\lambda\hat{B}}e^{\frac{\lambda}{2}\hat{A}} + \mc O(\lambda^3),
\end{equation*}
the quantum partition function can be expanded as
\begin{align*}
	&Z_\qim = \Tr\big( (e^{-\Delta\beta( \hat{H}^z + \hat{H}^x ) })^N\big)\\
	       &= \sum_\vS \prod_{n=1}^N\bra{\vS^n}e^{-\Delta\beta( \hat{H}^z + \hat{H}^x )} \ket{\vS^{n+1}}\\
	       &= \sum_\vS \prod_{n=1}^N\bra{\vS^n}e^{-\Delta\beta \hat{H}^z}e^{-\Delta\beta\hat{H}^x} \ket{\vS^{n+1}} +\mc O(\Delta\beta^3)\\ 
	       &= \sum_\vS \prod_{n=1}^N e^{-\Delta\beta H^z(\vS^n)}\prod_i \bra{\sigma^n_i}e^{\Delta\beta t\hS^x_i}\ket{\sigma^{n+1}_i}+\mc O(\Delta\beta^3)
\end{align*}
where $\vS^{N+1}\equiv \vS^1$. With
\begin{gather*}
	\bra{\sigma}e^{\Delta\beta t\hS^x}\ket{\sigma'}
	 = \cosh(\Delta\beta t)\delta_{\sigma,\sigma'}  +  \sinh(\Delta\beta t)\delta_{\sigma,-\sigma'}\\
	 \text{and}\quad
	 A e^{K^\tau \sigma\sigma'} 
	 = A \big(e^{K^\tau}\delta_{\sigma,\sigma'} + e^{-K^\tau}\delta_{\sigma,-\sigma'}\big)
\end{gather*}
we can identify
\begin{gather*}
	\bra{\sigma}e^{\Delta\beta t\hS^x}\ket{\sigma'} = A e^{K^\tau \sigma\sigma'},
	\,\,\text{where}\quad\\
	e^{-2K^\tau}=\tanh(\Delta\beta t)\quad\text{and}\quad
	A^2 = \sinh(2\Delta\beta t)/2.
\end{gather*}
Using this result and the definition $\mc{A}:=A^{LN}$ in the expansion of $Z_\qim$  ($N$ is the number of imaginary-time steps and $L$ the number of lattice sites), one obtains the connection between the quantum and the classical partition functions
\begin{align*}
	&\frac{Z_\qim}{\mc{A}} = \sum_\vS \prod_{n=1}^N e^{-\Delta\beta H^z(\vS^n) + \sum_i K^\tau_i \sigma_i^n\sigma_i^{n+1} } + \mc O(\Delta\beta^3)\\
	 &=  \sum_\vS e^{-\left(\sum_{n} K^z H^z(\vS^n)-\sum_{n,i} K^\tau_i \sigma_i^n\sigma_i^{n+1}\right)} + \mc O(\Delta\beta^3) \\
	 &=  Z_\cim + \mc O(\Delta\beta^3),
\end{align*}
with $K^z$ and $K^\tau$ as specified in Eq.\ \eqref{eq:param_connection}.
The normalization factor $\mc{A}$ cancels in the evaluation of expectation values for observables $\hat O=O(\{\hS^z_i\})$ that are diagonal in the $\{\hS^z_i\}$-eigenbasis and for which one obtains in the same way as for the partition functions
\begin{align*}
	\langle\hat{O}\rangle_\qim
	&= \frac{\Tr( e^{-\beta \hat{H}}\hat{O})}{Z_\qim} 
	 = \frac{\sum_\vS e^{-E_\cim(\vS)} O(\vS)}{Z_\cim} + \mc O(\Delta\beta^3)\\
	& = \langle O\rangle_\cim + \mc O(\Delta\beta^3).
\end{align*}
However, the factor $\mc{A}=[{\sinh(2\Delta\beta t)}/{2}]^{LN/2}$ [Eq.\ \eqref{eq:factor_A}] needs to be taken account of in the evaluation of non-diagonal observables such as the energy $\langle \hat{H}_\qim \rangle$ of the quantum system, as described in appendix~\ref{appx:energy}.

\section{Monte Carlo sampling and 1D cluster updates}\label{appx:MC}
With the quantum-classical mapping, described in section~\ref{sec:QIM_to_CIM}, we have constructed the classical model $E_\cim(\vS)$, Eq.~\eqref{eq:H_CIM}, in such a way that its partition function and expectation values of observables are identical to those of the quantum model as expressed in Eqs.\ \eqref{eq:equal_Z} and \eqref{eq:equal_obs}. The imaginary-time step $\Delta\beta=\beta/N$ of the quantum model enters the coupling constants $K^z$ and $K_i^\tau$ of the classical model according to Eq.\ \eqref{eq:param_connection} and the inverse temperature itself determines the number  $N$ of time slices, i.e., the extension of the \CIM{} in the time direction. The classical model is then formally sampled at $\beta_\cim=1$. In the Monte Carlo algorithm, we generate a Markov chain of classical states such that each state $\vS$ occurs with a frequency that corresponds to its weight $e^{-E_\cim(\vS)}/Z$ in the classical ensemble. As explained in section~\ref{sec:QIM_to_CIM}, expectation values of diagonal observables $\hat O=O(\{\hS^z_i\})$ can then be evaluated by averaging $O(\vS^n)$ (any choice of the time slice $n$ or additionally any average of the time slices $n$) with respect to the states generated by the algorithm. Non-diagonal observables can be addressed as exemplified in appendix~\ref{appx:energy}.

In Monte Carlo simulations, it is essential to obey \emph{detailed balance}, i.e., with the state probabilities $\pi(\vS):=e^{- E(\vS)}$ [in the following $E(\vS)\equiv E_\cim(\vS)$] and the state transition probabilities denoted by $p(\vS\to\vS')$, we require
\begin{equation}\label{eq:detailedBalance}
	\pi(\vS)p(\vS\to\vS')=\pi(\vS')p(\vS'\to\vS).
\end{equation}
Separating the transition probability into proposal and acceptance probabilities,
\begin{equation*}
	p(\vS\to\vS')=P(\vS\to\vS')A(\vS\to\vS'),
\end{equation*}
detailed balance can be achieved by using the \emph{Metropolis choice}
\begin{equation}\label{eq:acceptMetropolis}
	A(\vS\to\vS'):=\min\left(1,\frac{\pi(\vS')P(\vS'\to\vS)}{\pi(\vS)P(\vS\to\vS')}\right).
\end{equation}

As outlined in Sec.~\ref{sec:MC}, we base the simulation on flips of 1D clusters, oriented along the time direction, in order to avoid problematically low acceptance probabilities when decreasing $\Delta\beta$. This type of update is inspired by the Swendsen-Wang or Wolff cluster algorithms \cite{Swendsen1987-58,Wolff1989-62}.

The 1D cluster updates for the time direction of the \CIM{} \eqref{eq:H_CIM} are equivalent to cluster updates in an Ising chain $H_\eff= - K^\tau \sum_n \sigma^n_i \sigma^{n+1}_i + \sum_n h^n \sigma^n_i$ with site-dependent effective magnetic fields $h^n$ which encode the change in the number of flippable spins in time slices $n$. Denoting by $N_f^n$ the total number of flippable spins in time slice $n$ and by $\Delta N_f^n$ the change in this number due to flipping the spin $\sigma^n_i$, the effective magnetic field reads $h^n=K^z V\Delta N_f^n$. (Remember that the  potential term $\propto V$ in the Hamiltonian counts the number of flippable spins.) The chain consists of flippable spins and ends at time slices $m$ and $m'>m$ where the first non-flippable spins occur.

Because of the effective magnetic fields $h^n$, the actual Wolff cluster update \cite{Wolff1989-62} is not applicable (even for the 1D problem $H_\eff$). In the following, we describe an algorithm that is similar to the original Wolff cluster update in the sense that the clusters consist of parallel spins. Modifications are only due to the $h^n$. In principle, one can ignore the effective magnetic fields $h^n$ in the construction of the Wolff cluster. After the construction of a cluster, one would then flip it not with probability one as usual, but with a probability that takes the energy change $\Delta E_h:=K^z V \Delta N_f$ due to the effective fields $h^n$ and potential unflippable spins at the cluster ends into account. At least for small $|K^\tau/h^n|$, the resulting rejection rates would however be high. Also, the probability factor $e^{-\Delta E_h}$ may get small for big clusters even if $|K^\tau/h^n|$ is big and, thus, lead to a high rejection rate. Hence, it is favorable to take account of the energy changes due to the field terms $\propto h^n$ already during the construction of the clusters.
The algorithm works as follows:

(i) Start from a (consistent) random initial state $\vS_0$. Also, determine the number $N_f$ of flippable spins in $\vS_0$.

(ii) Choose a random flippable spin (site $i$, time slice $n$).

(iii) Let $\sigma_0:=\sigma^n_i$. Starting from the initial site $(n,i)$, go forward and backward along the direction of imaginary time, respectively, to build a 1D cluster of parallel spins. As long as the spin at the currently considered cluster boundary has magnetization $\sigma^{n'}_i=\sigma_0$ and is flippable, add it with probability
\begin{equation*}
	{q}(\Delta N_f^{n'}) := \left(1-e^{-2K^\tau}\right)\cdot\min\left(1,e^{-K^zV\Delta N_f^{n'}}\right)
\end{equation*}
to the cluster. In the following, let us denote the time slices that define the boundary of the obtained cluster by $m$ and $m'>m$, such that the cluster consists of time-slices $m+1,m+2,\dots,m'-1$. Let $f^m_i,f^{m'}_i\in\{0,1\}$ label whether the boundary spins are flippable (one) or not (zero).

(iv) Accept the flip of the cluster $\sigma_i^k\to {\sigma'}_i^k=-\sigma_i^k$ $\forall_{m<k<m'}$ with probability
\begin{multline}\label{eq:acceptWolff2}
	A(\vS\to\vS')=\min\Big(1,\frac{N_f}{N_f+\Delta N_f} e^{-K^zV\Delta N_f^{n}}\\
	 \times e^{-2K^\tau\sigma_0(\sigma_i^m+\sigma_i^{m'})}\,{\prod_{k=m,m'}\left[1-{q}(\Delta N_f^k)\right]^{-f^k_i \sigma_i^k \sigma_0}}  \Big).
\end{multline}
Why this rule guarantees detailed balance and is useful is explained below.

(v) If the number of cluster updates surpasses a certain threshold $\propto LN$, evaluate and store observables of interest, and reset the update counter to zero.

(vi) If you have accepted the transition in step (iv), update the spin configuration $\vS\to\vS'$ and $N_f\to N_f+\Delta N_f$. Go to step~(ii).

Eq.\ \eqref{eq:acceptWolff2} is based on the Metropolis choice \eqref{eq:acceptMetropolis} for the acceptance probability. The proposal probability for the cluster between time-slices $m$ and $m'$ is given by 
\begin{multline*}
	P(\vS\to\vS')=\frac{1}{N_f}\prod_{\stack{m<k<m'}{k\neq n}}q(\Delta N_f^k)\\
	\times \prod_{k=m,m'}\left[1-q(\Delta N_f^k)\right]^{f^k_i\delta(\sigma_i^k,\sigma_0)},
\end{multline*}
where $\delta(\sigma,\sigma')$ denotes the Kronecker delta.
Correspondingly,
\begin{multline*}
	P(\vS'\to\vS)=\frac{1}{N_f+\Delta N_f}\prod_{\stack{m<k<m'}{k\neq n}}{q}(-\Delta N_f^k)\\
	\times \prod_{k=m,m'}\left[1-{q}(\Delta N_f^k)\right]^{f^k_i\delta(\sigma_i^k,-\sigma_0)}.
\end{multline*}
Due to the fact that ${q}(-\Delta N_f^k)/{q}(\Delta N_f^k)=e^{K^zV\Delta N_f^k}$, we obtain
\begin{multline*}
	\frac{P(\vS'\to\vS)}{P(\vS\to\vS')} = \frac{N_f}{N_f+\Delta N_f}\,e^{\Delta E_h-K^zV\Delta N_f^n}\\ \times \prod_{k=m,m'}\left[1-{q}(\Delta N_f^k)\right]^{-f^k_i \sigma_i^k\sigma_0},
\end{multline*}
where $\Delta E_h=K^z V \sum_{k=m+1}^{m'-1}\Delta N^k_f=K^z V \Delta N_f$. Multiplying this with $\pi(\vS')/\pi(\vS)=e^{-\Delta E}$ with the total energy change $\Delta E=\Delta E_h+2K^\tau\sigma_0(\sigma_i^m+\sigma_i^{m'})$ yields Eq.\ \eqref{eq:acceptWolff2}. In the formula \eqref{eq:acceptWolff2} for the acceptance probability, one has only the factor $e^{-K^zv\Delta N_f^{n}}$ instead of $e^{-\Delta E_h}=e^{-K^z V\sum_{k=m+1}^{m'-1}\Delta N_f^{k}}$. So, the effective magnetic fields $h^n$ are taken into account during the cluster construction, and may reduce the cluster size, but they do not occur in the cluster flip acceptance formula and can hence not increase the rejection rate.

\section{Evaluation of the energy}\label{appx:energy}
The quantum Hamiltonian $\hat{H}\equiv\hat{H}_\qim$ [Eq.\ \eqref{eq:H_QIM}] is not diagonal in the $\{\hS^z_i\}$-eigenbasis and its expectation value can hence not be evaluated directly along the lines of Eq.\ \eqref{eq:equal_obs}. Based on the relation \eqref{eq:equal_Z} between the quantum and classical partition functions, an efficient way to evaluate the energy is to use that
\begin{align*}
	\langle\hat{H}\rangle_\qim
	&=\frac{1}{Z_\qim}\Tr\left(\hat{H} e^{-\beta \hat{H}}\right)
	 = \frac{-1}{Z_\qim}\partial_\beta Z_\qim\\
	&= \frac{-1}{N} \left(\frac{\partial_{\Delta\beta} Z_\cim}{Z_\cim}+ \frac{\partial_{\Delta\beta} \mc{A}}{\mc{A}}\right)
	   +\mc{O}(\Delta\beta^2)
\end{align*}
Using the relations \eqref{eq:param_connection} between the parameters of the \QDM{} and the \CIM{}, as well as 
$\mc{A}=A^{LN}=[{\sinh(2\Delta\beta t)}/{2}]^{LN/2}$ [Eq.\ \eqref{eq:factor_A}], one obtains
\begin{align*}
	&\langle\hat{H}\rangle_\qim
	= \frac{1}{N}\langle \partial_{\Delta\beta} E_\cim(\vS) \rangle_\cim - \frac{L}{A}\partial_{\Delta\beta} A +\mc{O}(\Delta\beta^2)\\
	&=\frac{1}{N} \sum_n \Big\langle \big( H^z(\vS^n) +\sum_i \frac{t\, \sigma_i^n\sigma_i^{n+1}}{\sinh(2\Delta\beta t)} \big) \Big\rangle_\cim\\
	&\phantom{=}\,\,\,- L t \coth(2\Delta\beta t) +\mc{O}(\Delta\beta^2)\\
	&= \sum_n\Big\langle \frac{1}{N}H^z(\vS^n) -\frac{1}{\beta}\sum_i\delta_{\sigma_i^n,-\sigma_i^{n+1}} \Big\rangle_\cim + \mc{O}(\Delta\beta).
\end{align*}
So what one basically needs to evaluate are averages of the number of flippable spins [$H^z(\vS^n)$] and the nearest-neighbor correlators $\sigma_i^n\sigma_i^{n+1}$ in the imaginary-time direction.
\begin{figure}[t]
\begin{centering}
\includegraphics[width=\columnwidth]{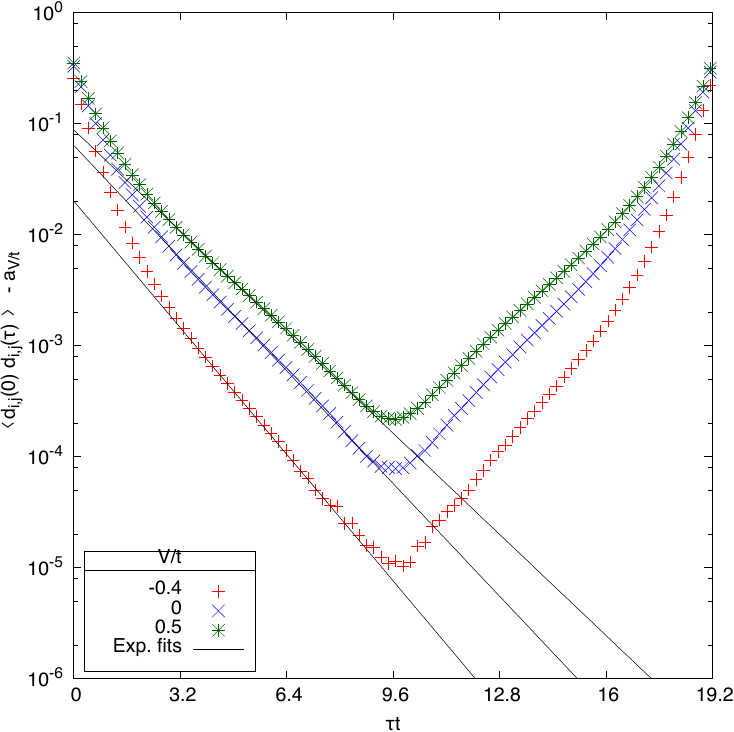}
\end{centering}
\caption{Determination of upper bounds on the energy gap by exponential fits of the dimer-dimer correlator $\langle\hat{d}_{i,j}(0)\hat{d}_{i,j}(\tau)\rangle_\qim$ [cf.\ Eq.~\eqref{eq:dimerdimer-correl}] for a system of $L=36\times 36$ sites, $\beta=19.2/t$, and $\Delta\beta=0.02/t$. In the figure, a fitted addend $a_{V/t}$ has been subtracted from the correlator.}\label{fig:corr_fit}
\end{figure}

\section{Evaluation of the energy gap}\label{appx:gap}
A common technique in quantum Monte Carlo simulations is to evaluate dynamical structure factors and spectral functions from imaginary-time correlation functions. This is complicated by the need of a, usually ill-conditioned, analytic continuation from the imaginary to real frequencies \cite{Gubernatis1991-44,Jarrell1996-269,Sandvik1998-57}. Away from the critical points the \QDM{} is expected to be in crystalline phases. In such phases, excitations are localized with little dispersion. Hence, the full spectral function is not very interesting and we can restrict the analysis to the estimation of excitation gaps -- in particular to assert the gapped nature of the plaquette phase. This can be done without the need for analytic continuation. See, for example, Refs.\ \cite{Assaad1996-65,Assaad1999-83,Ivanov2004-70,Ralko2006-74} for similar studies.

We want to estimate the energy gap to excited states by evaluating imaginary-time correlation functions
\begin{align*}
	\langle \hat A(0)\hat A^\dag(\tau)\rangle
	&= \frac{1}{Z}  \Tr \left( \hat A e^{-\tau \hat H} \hat A^\dag e^{-(\beta-\tau)\hat H} \right).
\end{align*}
If $\tau$ and $\beta-\tau$ are both big enough in comparison to the gap to the second excited state, one can expect the correlation functions to have a $\cosh$ form. For a generic operator $\hat A=\sum_{ij} a_{ij}\ket{i}\bra{j}$, with the eigenstates $\ket{i}$ ($i\in \mathbb{N}_0$) of the system ordered according to increasing energies $E_i$ and gaps denoted by $\Delta E_{j,i}:=E_j-E_i$, one gets
\begin{align*}
	&\langle \hat A(0)\hat A^\dag(\tau)\rangle \\
	&= \frac{1}{2Z}\sum_{ij} |a_{ij}|^2 ( e^{-\tau  E_j}e^{-(\beta-\tau) E_i} + e^{-\tau E_i}e^{-(\beta-\tau) E_j} )\\
	&= \frac{1}{2Z}\sum_{ij} |a_{ij}|^2 e^{-\beta E_i} ( e^{-\tau \Delta E_{j,i}} + e^{-(\beta-\tau)\Delta E_{j,i}} )\\
	&= \frac{1}{Z}\sum_{ij} |a_{ij}|^2 e^{-\beta (E_j+E_i)/2}  \cosh( (\beta/2-\tau)  \Delta E_{j,i} ),
\end{align*}
i.e., a sum of $\cosh$ terms with non-negative coefficients that decay exponentially in $\beta$ and $E_j+E_i$ (due to the normalization factor $1/Z$ rather in $E_j+E_i-2E_0=\Delta E_{j,0}+\Delta E_{i,0}$). The ``saturation'' value $\langle \hat A(0)\hat A^\dag(\beta/2)\rangle = \frac{1}{Z}\sum_{ij} |a_{ij}|^2 e^{-\beta (E_j+E_i)/2}$ of the correlator ($\tau=\beta/2$) has for low temperatures $\beta\Delta E_{1,0}\gg 1$ the value $|\langle \hat A\rangle_{gs}|^2$. As exemplified in Fig.~\ref{fig:corr_fit}, one can hence extract the gap of the system by fitting a few leading terms of the sum to the imaginary-time correlation functions, the simplest expression being $a+b\cdot\cosh( (\beta/2-\tau)  \Delta E _{1,0})$. To this purpose, we choose the dimer-dimer correlator
\begin{equation}\label{eq:dimerdimer-correl}
	\langle\hat{d}_{i,j}(0)\hat{d}_{i,j}(\tau)\rangle_\qim
	=\frac{1}{4}\langle (\sigma_i^n\sigma_j^n+1)(\sigma_i^{n'}\sigma_j^{n'}+1)\rangle_\cim
\end{equation}
and the plaquette flip correlator
\begin{equation*}
	\langle\hS^x_i(0)\hS^x_i(\tau)\rangle_\qim
	=\frac{1}{(\Delta\beta t)^2}\langle \delta_{\sigma_i^n,-\sigma_i^{n+1}}\delta_{\sigma_i^{n'},-\sigma_i^{n'+1}}\rangle_\cim
\end{equation*}
where $\hat{d}_{i,j}:=(\hS^z_i\hS^z_j+1)/2$ is the dimer operator for edge $(i,j)$, integers $n\in[1,N]$ label imaginary-time slices, and $n'=n+\tau/\Delta\beta$.

As described in section~\ref{sec:QDM_to_QIM}, the dimer-model Hilbert space corresponds to the symmetric sector of the Ising model on the dual lattice, spanned by states $\ket{\vS}+\ket{-\vS}$ with ground states $\vS$ of the classical Ising model. For the gap estimation, it is hence important to choose operators $\hat A$ that do not connect the symmetric and anti-symmetric subspaces. For example, $\hat A=\hS_i^z$ transforms spin-flip symmetric into anti-symmetric states. The analysis of the corresponding imaginary-time correlator would hence yield an estimate for the gap between the ground-state energies of the two sectors. In contrast, the dimer operator $\hat A=\hat{d}_{i,j}$ and the plaquette-flip operator $\hat A=\hS^x_{i}$ are block-diagonal and yield an estimate for the gap to the first excited state in the symmetric subspace.

\section{Perturbation theory for the star phase}\label{appx:pert}
The ideal star state \eqref{eq:starState} is a product state with 3-plaquettes on two of the three triangular sublattices (say $A$ and $B$) and 0-plaquettes on $C$. It is the ground state for $V/t \rightarrow -\infty$, where the potential energy term selects the classical dimer coverings with the maximum number of $3$-plaquettes. For a perturbative analysis in $\lambda:=t/V$, we write the Hamiltonian \eqref{eq:H_QDM} in the form
\begin{equation}\label{eq:H_QDM_rewrite}
	\hat{H}_\qdm = V\big(-\lambda\sum_i\hat{f}_i + \hat{N}_3\big),
\end{equation}
where $\hat{f}_i = \left(\ketbra{\pUp_i}{\pDown_i}+h.c.\right)$ flips plaquette $i$, and $\hat{N}_3 = \sum_i \left(\ketbra{\pUp_i}{\pUp_i}+\ketbra{\pDown_i}{\pDown_i}\right)$ counts the total number of flippable plaquettes.

Let us denote the energy of the $i^\text{th}$ unperturbed eigenstate by $E^{(0)}_i$ and $\ket{\psi_0}:=\ket{\psi_\Star}$. For $\lambda=0$, the first excited states $\ket{\psi_{1,i}} := \hat{f}_i\ket{\psi_0}$ are obtained by flipping single plaquettes. The other two degenerate ground states can be disregarded for the following as they can only be reached by an extensive number of flips. Up to second order, the perturbed energy is
\begin{equation}
	\frac{E_\Star^{(2)}}{V} = \frac{E^{(0)}_0}{V}+\lambda^2 \sum_i\frac{|\bra{\psi_{1,i}}\hat{f}_i\ket{\psi_0}|^2}{E^{(0)}_0/V-E^{(0)}_1/V} + O(\lambda^3),
\end{equation}
since the linear term  $\bra{\psi_0}\hat{f}_i\ket{\psi_0}$ is zero. Applying Eq.~\eqref{eq:H_QDM_rewrite}, we find
\begin{align}
	E_\Star^{(2)} &\quad\,=\quad\frac{2L}{3}V+\lambda^2 V \frac{2L}{3}\frac{1}{\frac{2L}{3}-\left(\frac{2L}{3}-3\right)}\nonumber\\
	          &\overset{(\lambda = t/V)}{=} L\cdot\left(\frac{2V}{3}+\frac{2t^2}{9V}\right).
	\label{eq:H_Star}
\end{align}

\end{document}